# Trustworthy Mobile Edge Caching: A Blockchain Approach to Mitigate Malicious Nodes and Incentivize Cache Sharing

**Motahare Ebrahimi[1], Nastooh Taheri Javan[2*], Seyedakbar Mostafavi[3*] and Fatemeh Pakzaban[4]**
[1,3,4] Department of Computer Engineering, Yazd University, Yazd 89158-18411, Iran
[2] Computer Engineering Department, Imam Khomeini International University, Qazvin 34148-96818, Iran
[1]motahare.ebrahimi@stu.yazd.ac.ir , [2]nastooh@eng.ikiu.ac.ir, [3]a.mostafavi@yazd.ac.ir, [4]f.pakzaban@stu.yazd.ac.ir

[*]Corresponding authors: Nastooh Taheri Javan and Seyedakbar Mostafavi

**ABSTRACT** As mobile network traffic continues to grow, content caching on edge servers is critical for reducing latency. However, challenges such as malicious edge servers that may delete or manipulate cached content, along with the limited capacity of these servers, need to be addressed. To overcome the capacity limitations, helper mobile nodes can contribute their cache resources. However, due to their selfish behavior, an incentive mechanism is necessary to encourage resource sharing. Additionally, these helper nodes can also be malicious. This paper proposes a blockchain-based trust management mechanism that addresses these challenges by accurately identifying trustworthy edge servers and mobile nodes. The proposed mechanism calculates both direct and indirect trust using smart contracts, ensuring that malicious nodes are effectively filtered out. Trustworthiness is determined based on mobile node satisfaction with the quality of service, and trust data is securely stored on the blockchain. To combat node selfishness, a reward mechanism is introduced to incentivize cache sharing. Furthermore, a blockchain-based authentication mechanism protects against node impersonation. Our approach optimizes trust, cache capacity, and cost efficiency while considering mobile node mobility, energy consumption, and computational power constraints during the consensus process. Simulation results show that the proposed method can accurately distinguish between honest and malicious servers, even with a 10% noise in data.

## I. INTRODUCTION

With the rapid adoption of smartphones and mobile applications, the number of mobile users in society has grown exponentially. This has led to an increased demand for diverse content delivery. However, mobile users face challenges due to limitations in resources such as bandwidth and energy. As a result, many users struggle to obtain the desired quality of service (QoS), and the competition for these limited resources often results in delays in content delivery [1].

Cloud computing, which centralizes services such as data storage, computation, and processing, has been an essential solution to provide scalable services [2]. Despite its benefits, centralized cloud computing introduces several challenges, such as a single point of failure, poor scalability, and network congestion. These issues arise due to the centralization of data storage, which can overwhelm the system when there is a surge in user demand. Moreover, physical distance from cloud data centers increases network latency, leading to slower responses and reduced QoS [3, 4]. Additionally, privacy concerns have made cloud data storage impractical in certain scenarios [5, 6].

Edge computing has emerged as a solution to overcome these limitations while retaining the advantages of cloud computing [7]. By deploying edge servers closer to mobile users, data is cached locally, which reduces the traffic between mobile devices and cloud data centers. This not only enhances content delivery performance but also reduces latency and alleviates congestion during peak times [8, 9]. Edge caching improves system efficiency by pre-storing popular content, making it available for faster retrieval, and reducing the need to fetch content from remote servers [10].

However, the deployment of edge caching systems is not without challenges. Edge servers, typically owned by third-party entities, can be compromised by malicious actors who manipulate cached content to maximize personal gain. This manipulation could result in the delivery of fake content, exposing users to malicious data or preventing access to legitimate content [1]. As a result, ensuring the reliability of edge nodes is critical to prevent content tampering and ensure the system's integrity. Trust management in edge computing, therefore, plays a crucial role in building confidence in edge infrastructures and maintaining secure relationships between mobile users and edge servers [11].

The management of trust in edge computing can be classified into centralized, semi-centralized, and distributed systems. While centralized systems rely on a single authority to manage trust, they suffer from scalability issues and risks associated with a single point of failure. Semi-centralized and distributed systems have been proposed to address these issues but still face challenges in terms of reliability, data consistency, and auditability, particularly in the context of the Internet of Things (IoT) [12, 13].

Blockchain technology has emerged as a decentralized, tamper-resistant solution for improving trust management systems. It provides the transparency, stability, and resistance to manipulation required for ensuring secure and reliable interactions between edge servers and mobile devices [14]. By leveraging blockchain, trust management can be decentralized, enabling secure and auditable transactions without the need for a central authority. Blockchain offers significant advantages for trust evaluation, especially in systems where malicious behavior can be a concern.

To address the challenges of content manipulation by malicious servers, this paper proposes a blockchain-based trust management system. In this system, mobile nodes continuously assess the trustworthiness of edge servers based on the quality of content delivery, ensuring that only reliable edge servers are used for caching and data storage. A private blockchain records the transactions between mobile users and edge servers, with the blockchain serving as a transparent and immutable log for trust evaluation. The system uses a two-stage consensus mechanism to ensure trustworthiness and smart contracts to calculate and store trust values.

Furthermore, mobile nodes assist in content delivery by serving as "caching mobile users (CMUs)," contributing cache capacity to improve content availability. However, this introduces additional challenges such as node mobility, selfish behavior, malicious activity, and resource limitations (e.g., energy and computational power). This paper addresses these issues by proposing an incentive mechanism to encourage mobile nodes to share cache resources, a trust evaluation model that mitigates the effects of collusion, and an authentication mechanism to prevent identity forgery and attacks like Whitewashing.

The main contributions of this paper are:

- Trust evaluation for edge servers based on the quality of caching and protection against mobile node collusion;
- An incentive mechanism for mobile nodes to share cache capacity;
- An authentication mechanism to prevent identity forgery and protect against attacks like Whitewashing;
- A two-phase consensus mechanism, designed with consideration for the mobility of mobile nodes, to address their energy and computational limitations by delegating the consensus process exclusively to high-performance edge servers—thereby reducing both latency and operational costs.

The structure of this paper is organized as follows: Section II reviews related work in the field. Section III introduces the network model, and Section IV details the proposed method. Section V presents the evaluation of the proposed system, and Section VI concludes the paper.

## II. Related Works

In recent years, a significant body of research has been dedicated to improving content delivery in mobile networks, particularly in the context of cloud and edge computing. While cloud computing offers scalable solutions, it faces challenges such as latency, bandwidth limitations, and privacy concerns. Edge computing, with its decentralized approach, has emerged as a promising solution to these issues by bringing computation and data storage closer to the user. However, trust management remains a critical challenge, especially with the potential for malicious activity in edge nodes. Blockchain technology has been increasingly explored as a means to secure and decentralize trust management in such systems, providing a transparent and tamper-resistant infrastructure for content delivery.

In this section, recent studies related to this research are briefly reviewed. Among these, the studies that are most closely related to our proposed approach have been briefly compared.

Xu et al. [10] introduced a reliable blockchain-based edge caching scheme for mobile users in Mobile Cyber-Physical Systems (MCPS). Their approach leverages blockchain to monitor caching transactions between edge nodes and mobile users in a distributed manner, preventing any alterations to cache service information. Additionally, they developed a trust management mechanism for selecting trustworthy edge nodes that can provide high-quality caching services to mobile users within social groups. While this work enhances reliability, our approach goes further by incorporating a two-stage consensus mechanism to ensure both security and efficiency in trust evaluation for diverse mobile users.

Zhang et al. [15] proposed a blockchain-based trust management system for the Internet of Vehicles (IoV) that can identify and penalize vehicles that send malicious messages, preventing tampering with reputation data. This approach primarily addresses security in vehicular networks, whereas our work focuses on the edge computing domain and extends beyond security by implementing a cooperative-node selection mechanism that incentivizes mobile nodes to share cache resources while ensuring trustworthiness.

Li et al. [16] developed a trust evaluation scheme for vehicles and roadside units (RSUs) aimed at securing content delivery in vehicular networks. They used an incentive scheme to encourage trust improvements in content delivery. Our work draws on a similar incentive concept but adapts it for edge caching in mobile networks, rewarding nodes based on trust and shared resources to improve caching performance and security in edge environments.

Xiao et al. [17] introduced a blockchain-based trust mechanism to enhance computational performance in Mobile Edge Computing (MEC) by discouraging selfish attacks and service history forgeries. While similar in concept, our approach expands on this by incorporating a structured trust evaluation that distinguishes between malicious, low-quality, and high-quality edge servers, allowing for more nuanced service evaluations in our caching network.

Debe et al. [18] proposed a decentralized trust system for public fog nodes, using blockchain to address the single point of failure in centralized architectures. By recording and analyzing past interactions, they build trust between IoT devices and fog nodes. Although this approach builds reliability in public fog nodes, our model applies a targeted trust management strategy for caching interactions in mobile edge networks, emphasizing real-time trust evaluations.

Pathak et al. [19] introduced a trust-based access control mechanism for Edge-IoT networks. Their system dynamically evaluates trust to identify and isolate malicious IoT devices. In contrast, our work integrates access control with trust management in a way that facilitates cache resource distribution by edge servers, utilizing caching mobile users (CMUs) to enhance content accessibility even under resource constraints.

Zarandi et al. [20] proposed a blockchain-based trust management system for the Social Internet of Things (SIoT), relying on trust evaluations based on reputation and social relationships. While their system achieves consistency in trust assessment, our model applies a similar reputation-based approach to improve the caching process, selecting caching mobile users (CMUs) based on trust values that directly impact the reliability of content delivery.

Wang et al. [21] presented a computational trust management system for IoT in smart cities, which improves quality of service by filtering participants based on trust values. Although effective in smart city IoT, our work differs by focusing on trust-based caching and reward mechanisms specifically designed for mobile nodes, optimizing cache utilization and trust management within edge computing.

Deng et al. [22] developed a three-layer trust evaluation system for MEC networks that assesses edge servers based on identity, capability, and behavior trust. Our proposed model, while similar in incorporating multiple dimensions of trust, introduces a two-stage consensus for trust and caching performance, which is particularly suited for the dynamic mobile edge environment.

Cheng et al. [23] proposed a resource allocation system for IoT using blockchain, which incorporates a reputation-based mechanism to assign high-reputation servers to end users. While their reputation-based allocation addresses IoT service quality, our system also considers mobile node mobility, adapting content caching for constantly moving mobile devices.

Shen et al. [24] introduced a blockchain-based trust and content dissemination system for Vehicle-to-Vehicle (V2V) networks, with reputation scores to enable secure video content sharing. Though focused on V2V, this study's trust management mechanism is analogous to ours in that we both emphasize fair and secure content sharing. However, our approach directly incentivizes caching participation among caching mobile users (CMUs), specifically designed for mobile networks.

Zihao et al. [25] proposed a collaborative content delivery mechanism using Parking Vehicles (PVs) for secure and efficient content delivery. Their model improves delivery by evaluating PVs' honesty based on reputation. Our work extends this idea of honesty-based collaboration to mobile nodes in edge caching, rewarding them based on their trustworthiness and cache contributions, thereby enhancing reliability in mobile edge networks.

By comparing and reviewing the studies conducted in Table I, none have fully addressed the challenges of trust management in edge computing. These challenges include external attacks such as user impersonation in edge computing, Whitewashing attacks, trust manipulation and the trust calculation algorithm by malicious nodes. Additionally, using an appropriate trust mechanism to prevent collusion among malicious nodes that could increase the trust of malicious edge servers and decrease the trust of well-behaved edge servers has not been addressed. In this article, a blockchain-based authentication method is proposed to counter user impersonation, along with a blockchain-based trust management mechanism to prevent trust manipulation and tampering with the trust calculation algorithm, thereby addressing the

afore0mentioned challenges. Furthermore, the mobility of mobile nodes and their limitations in energy consumption and computational power have been considered.

Recent studies such as [26] and [27] have proposed cooperative caching schemes that consider trust and incentive mechanisms. The work in [26] utilizes reputation-aware clustering in vehicular networks, while [27] explores social trust for content placement in mobile networks. Our proposed model differs by leveraging a blockchain-based architecture with decentralized trust calculations and a two-stage consensus protocol, ensuring enhanced resilience against malicious nodes and incentivized cooperation among mobile users.

Yao et al. [28] developed a task offloading approach using multi-agent reinforcement learning in crowd-edge environments, where devices learn to offload tasks efficiently to reduce delay and energy use. Liu et al. [29] introduced DKGAuth, a blockchain-based method for distributed key generation and cross-domain authentication in IoT networks, removing the need for a central authority. Both works highlight the role of intelligent and decentralized solutions in improving efficiency and security in edge and IoT systems.

The approaches reviewed in this section are summarized in Table 1. For this purpose, the strengths and weaknesses of previous studies, in relation to our proposed approach, are listed in the table.

## III. System Model

To ensure scalability in environments with numerous mobile nodes and edge servers, the system distributes decision-making to the edge layer. Trust evaluations and caching allocations are computed locally using lightweight methods, minimizing centralized overhead. Additionally, the reward mechanism is based on localized trust assessments and node availability. In future work, we plan to introduce cluster-based trust regions and adaptive delegation to further improve performance in dense deployments.

In this section, the system model and network assumptions considered in this research are briefly explained. To ensure optimal performance and security in content delivery, the proposed system utilizes a hybrid architecture that simultaneously considers trust and optimal resource allocation. Given the challenges faced in edge computing networks and the increasing demand for efficient content delivery, the proposed model leverages blockchain technology for trust evaluation and resource management, ensuring transparency and security in the processes. This section details the proposed architecture and clarifies how it addresses scalability, service quality, and security challenges in edge computing systems.

As illustrated in Figure 1, the proposed architecture comprises a content provider, a registration center (RC), an authentication server (AS) located in the cloud, multiple edge servers (ES), and several mobile devices (MU). Each edge server, along with its connected mobile devices, forms a group, with the edge server serving as the group manager.

TABLE 1
THE ADVANTAGES AND DISADVANTAGES OF TRUST MANAGEMENT METHODS IN EDGE COMPUTING

| Ref. | Advantages | Disadvantages |
|---|---|---|
| [10] | • Layered coding mechanism for reduced content delivery delay<br>• Max-min fairness algorithm for cache allocation<br>• Blockchain for storing user-edge interactions, prices, and trust levels | • Ignores mobility and collusion among nodes<br>• Vulnerable to ballot stuffing, bad-mouthing, and Whitewashing attacks |
| [14] | • Hybrid consensus (PoS and PoW) for faster vehicle reputation updates<br>• Blockchain storage for data and signature security | • Vulnerable to Whitewashing attacks |
| [15] | • Bargaining game model enhances trust and security in content delivery | • Susceptible to Whitewashing<br>• Risk of trust manipulation |
| [16] | • Reduces selfish edge attacks, response delay, and energy consumption | • Vulnerable to various attacks, including bad-mouthing and Whitewashing |
| [17] | • Calculates service-specific reputations<br>• Prevents collusion in reputation assessments | • Vulnerable to impersonation and Whitewashing |
| [18] | • Resists DoS, self-promotion, Whitewashing, Sybil, and bad-mouthing attacks | • Limited security measures for malicious entities |
| [19] | • Lightweight reputation calculation via information entropy<br>• Resists bad-mouthing, ballot stuffing, DoS, and storage attacks | • Lacks protection against attacks on fog nodes |
| [20] | • Defends against bad-mouthing, stuffing, and selective forwarding | • Insufficient mechanisms for collusion and complex fraud |
| [21] | • Resistant to Whitewashing and collusion<br>• Reputation growth only through sustained high-quality services | • Potential for trust manipulation |
| [22] | • End-user credibility evaluation for reputation to prevent collusion<br>• Corrective parameter limits reputation manipulation | • Vulnerable to impersonation and Whitewashing |
| [23] | • Decentralized V2V content dissemination, resistant to content manipulation and DDoS | • Risk of collusion among client vehicles |
| [24] | • Incentives for honest nodes improve trust and security in content delivery | • Potential for trust manipulation |

In this architecture, the interaction between edge servers and mobile devices, including trust evaluation and authentication processes, plays a key role in the system's performance. The proposed model in this study, through precise assessment of the service quality provided by edge servers and mobile devices' participation in resource sharing, contributes to creating a collaborative environment. Additionally, by using mechanisms like blockchain and consensus algorithms, the system prevents malicious behaviors and ensures that content delivery processes are secure and trustworthy.

Each MU selects the optimal ES from the available servers in its vicinity based on two criteria: 1) the ES's trustworthiness and 2) the ES's caching price, which should be low, and its capacity, which should be high [10]. After selecting an ES, the mobile node undergoes authentication by the ES and establishes a connection. The mobile node then generates a caching token to safeguard against man-in-the-middle attacks and data manipulation, which is sent to the ES. The ES verifies the caching token, and if it is valid, provides the requested service to the MU. Based on the MU's satisfaction level with the quality of ES services, the trustworthiness of the ES is calculated and stored on the blockchain. Specifically, using the satisfaction value provided by the user, the trust level of the ES is computed via a smart contract. Since the storage and calculation of trust occur within the blockchain, manipulating the trust values or the trust calculation algorithm is prevented.

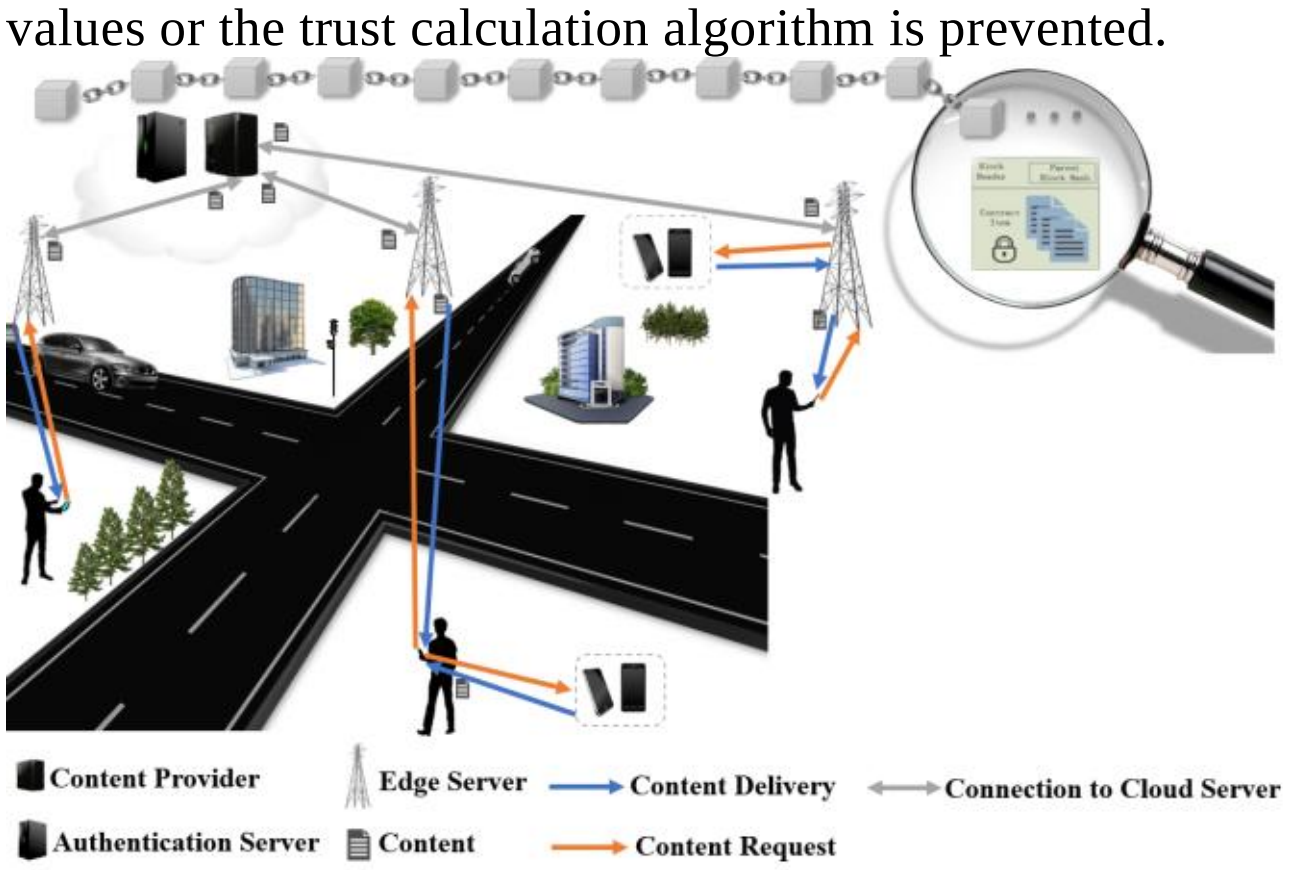

FIGURE 1. Content Delivery System Model in Mobile Networks

The distribution of caching resources by the ES follows the Max-min fairness algorithm. If the total demand from MUs is less than the ES's caching capacity, each MU receives the required caching resources. If the demand exceeds the ES's capacity, each MU will receive only a portion of the required caching resources [10]. When the ES's cache capacity is fully utilized, the ES solicits nearby mobile nodes to share their cache capacity. Each ES selects caching mobile users (CMUs) based on two criteria: trustworthiness and high sharing capacity. The ES then rewards caching mobile users (CMUs) according to the amount of cache capacity shared and their trust level.

In the proposed approach, the concern of mobile nodes regarding the rejection of cache service requests by edge servers is greatly alleviated due to the use of the Max-min fairness algorithm in the distribution of edge server cache resources and the use of the cooperative mobile node's cache capacity. After the cache service is received by the mobile node, the trust, price, and remaining capacity of the edge server are calculated and recorded on the blockchain using the edge server's public key. The blockchain utilized in this paper is a private blockchain.

Additionally, the ESs are authenticated by the AS, and the RC employs the AES algorithm to generate public and private keys for the ESs. Given that honest edge servers may occasionally underperform by providing low-quality services, and that malicious edge servers might at times deliver desirable services, we include low-quality edge servers in the classification scheme of our proposed approach. Accordingly, in our method, edge servers are categorized into three groups: malicious, low-quality, and high-quality. This classification is crucial, as it enables us to accurately identify samples corresponding to each category within our approach. To this end, we will utilize a confusion matrix in the evaluation section to accurately identify samples associated with each category, aiming to achieve the highest possible accuracy and precision in their categorization and detection.

## IV. Proposed Method

In this section, the steps of the proposed method are explained in detail. Figure 2 summarizes the main steps of the proposed approach. The following sections elaborate on the authentication process, consensus mechanism, content model, cache token creation mechanism, pricing, trust evaluation, cache resource allocation for ESs, and rewards for cooperative mobile nodes within the proposed architecture. Furthermore, the optimal cache demand of MUs from the ES is calculated.

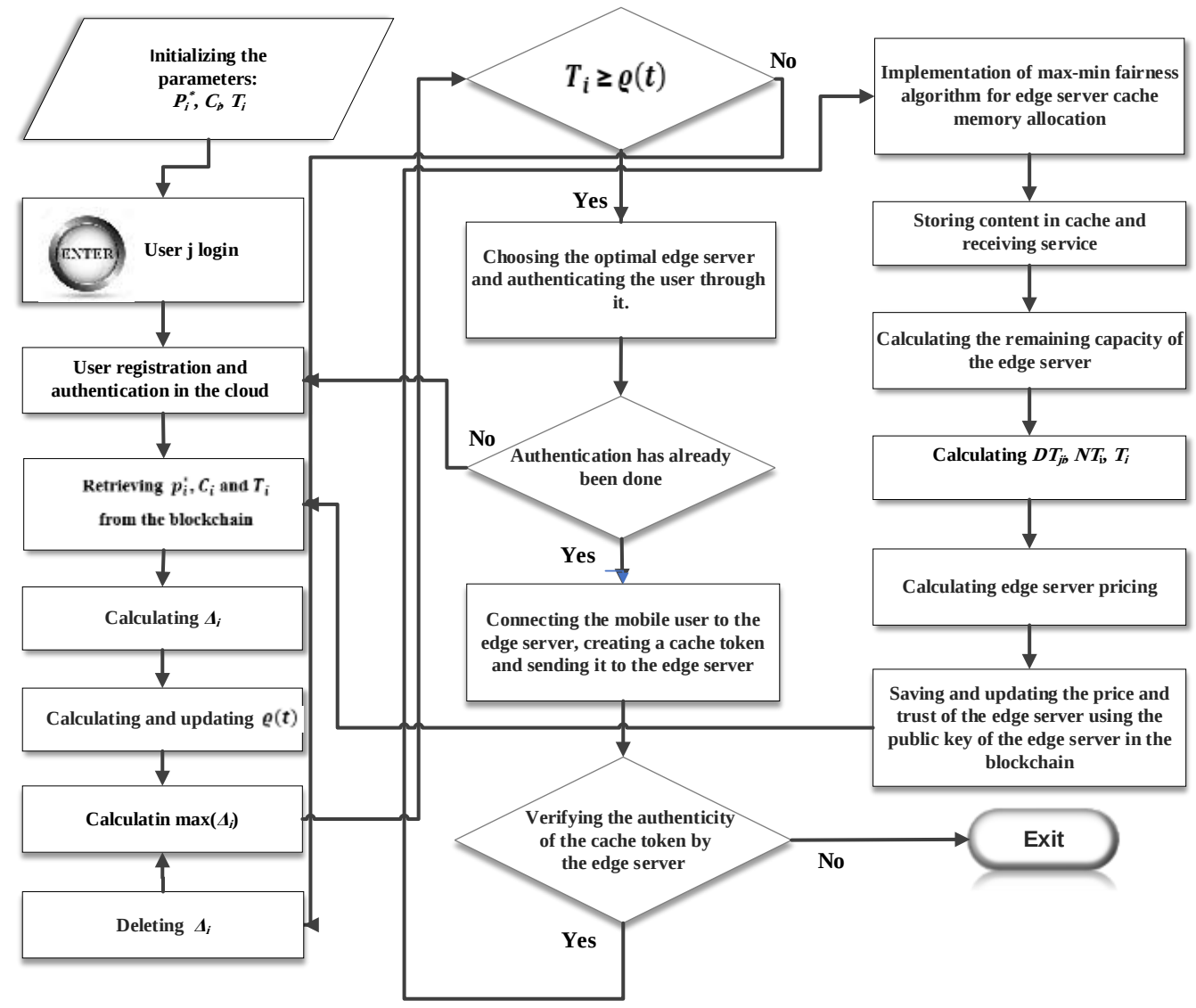

FIGURE 2. Flowchart of the Proposed Method

### A. Authentication process

Some mobile users (MUs) may act maliciously and jeopardize network security by using fake identities. For example, these MUs might collude with other nodes to either increase or decrease the trust in a malicious edge server (ES) or cooperative node, or they could cache virus-infected content in the ES. Such actions pose significant security threats and compromise overall network integrity. To counter these threats, an authentication mechanism similar to the one described in [30] is proposed. In this mechanism, blockchain is used to store the required information in the authentication process to address the time-consuming issue of authenticating mobile nodes when changing groups.

The proposed method is as follows: MUs first register with the cloud-based Registration Center (RC), which uses the RSA algorithm to generate and securely distribute public ($pk_j$) and private ($sk_j$) keys to the MUs for authentication. After registration, MUs must authenticate with the cloud-based Authentication Server (AS), which has access to the keys stored in the RC. The authentication process between the AS and MU is illustrated in Figure 3.

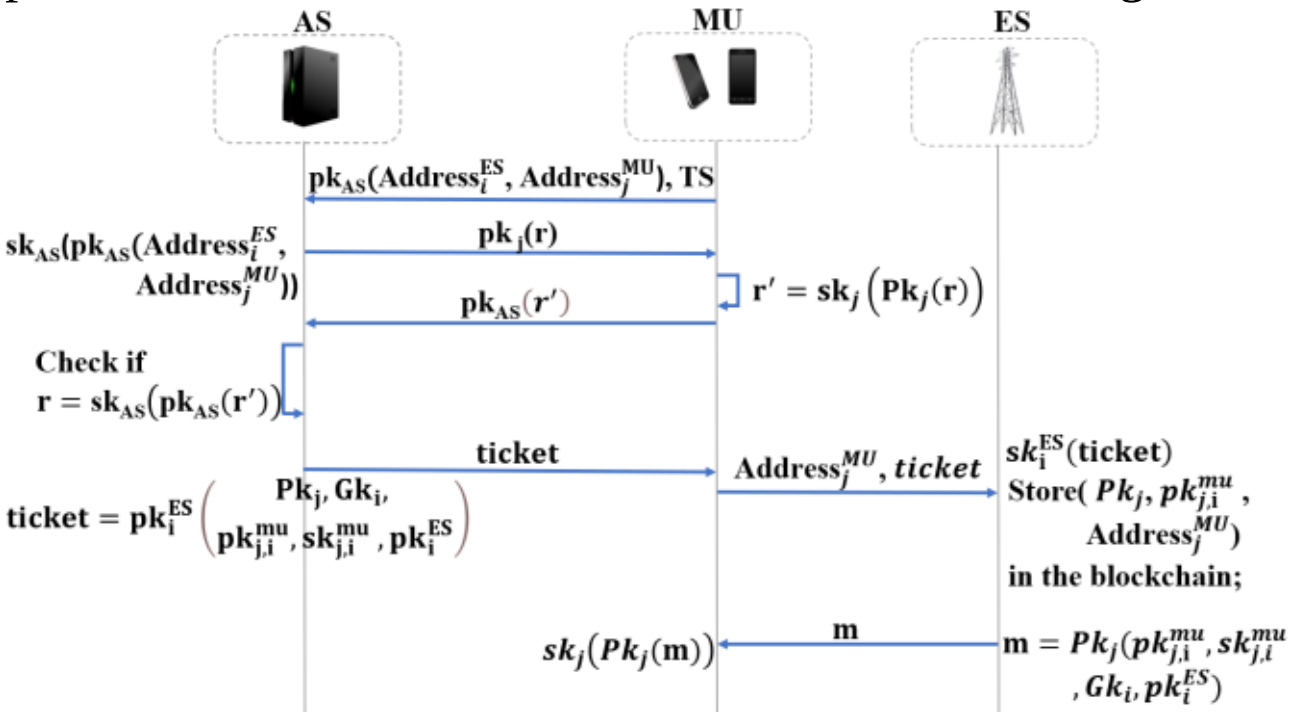

FIGURE 3. **The Authentication Mechanism Between MU and AS**

The MU encrypts a message containing its address and the address of the ES it wishes to connect to using the AS's public key ($pk_{AS}$) and sends it, along with a timestamp (TS) to the AS. The timestamp is included to prevent replay attacks. The AS decrypts the message with its private key ($sk_{AS}$), retrieves the MU's public key information from the RC, and generates a random number $r$. It then sends an encrypted message $pk_j(r)$ to the MU. Upon receiving this message, the MU decrypts it using its private key ($sk_j$) to obtain the random number $r$, then encrypts $r$ with the AS's public key ($pk_{AS}$) and sends it back to the AS. The AS decrypts the response with its private key ($sk_{AS}$) and verifies whether the random number matches the one it initially generated. If they match, the MU is authenticated.

The AS then sends a ticket to the MU containing the following information, all encrypted with the ES's public key ($pk_i^{ES}$): $pk_j$, the group public key $GK_i$ of group $i$, $pk_{j,i}^{mu}$ (the MU's public key in group $i$), $sk_{j,i}^{mu}$ (the MU's private key in group $i$), and $pk_i^{ES}$. After receiving the ticket, the MU initiates a connection to the ES, sending its address and the ticket to the ES. The ES decrypts the ticket using its private key ($sk_i^{ES}$), stores $pk_{j,i}^{mu}$ and the MU's address on the blockchain, and sends a message $m$ containing $m = pk_j(pk_{j,i}^{mu}, sk_{j,i}^{mu}, Gk_i, pk_i^{ES})$ to the MU. The MU decrypts this message using its private key ($sk_j$), retrieves the content, and stores it.

If a mobile user (MU) moves to a new edge server (ES) and thus joins a new group, it is not required to re-authenticate with the Authentication Server (AS) due to the high session delay and time-consuming authentication steps. Instead, each time the MU initially connects to an ES, the ES records the MU's address, $pk_{j,i}^{mu}$, and public key on the blockchain. When the MU connects to a new ES, this new ES can simply query the blockchain for the MU's address.

As illustrated in Figure 4, the MU previously connected to server $ES_i$ sends a message $m_1$ containing $m_1 = (Address_j^{MU}, TS, sk_j(Address_j^{MU}, TS))$ to $ES_{i+1}$. $ES_{i+1}$ retrieves the MU's public key from the blockchain to verify its authenticity. If the signature is valid, $ES_{i+1}$ responds with a message $m_2$ containing $m_2 = (Address_{i+1}^{ES}, TS+1, pk_j(Address_{i+1}^{ES}, TS+1, GK_{i+1}, pk_{j,i+1}^{mu}, sk_{j,i+1}^{mu}, pk_{i+1}^{ES})$. The MU then uses its private key to access all the keys and information in $m_2$. To confirm that it has acquired the new group's keys, the MU sends a message $m_3$ containing $m_3 = (Address_j^{MU}, TS+2, pk_{i+1}^{ES}(Address_j^{MU}, TS+2))$ to $ES_{i+1}$. $ES_{i+1}$ decrypts $m_3$ using its private key ($sk_{i+1}^{ES}$) and verifies the message. Upon successful verification, the MU is authenticated with $ES_{i+1}$, becomes a member of the new group, and establishes a connection with the server. Finally, $ES_{i+1}$ in group $i$+1 generates a new block of MU keys, including $pk_j$, $pk_{j,i+1}^{mu}$, $Address_j^{MU}$, updates the old block, and sends a transaction for verification across the blockchain network.

### *B. Consensus Mechanism*

The proposed framework limits consensus participation to edge servers due to their superior stability and resource capacity compared to mobile users. This design prevents frequent reconfiguration in response to node mobility. For large-scale networks, consensus can be further optimized using cluster-based delegation or hierarchical coordination among edge servers. This approach preserves efficiency while accommodating high scalability requirements.

In the proposed approach, we utilize a private blockchain with a consensus mechanism inspired by [31]. Given that mobile users (MUs) have limited computational power and energy, the approach incorporates lightweight nodes in the blockchain that do not participate in the consensus process. Only edge servers (ESs) are involved in the consensus.

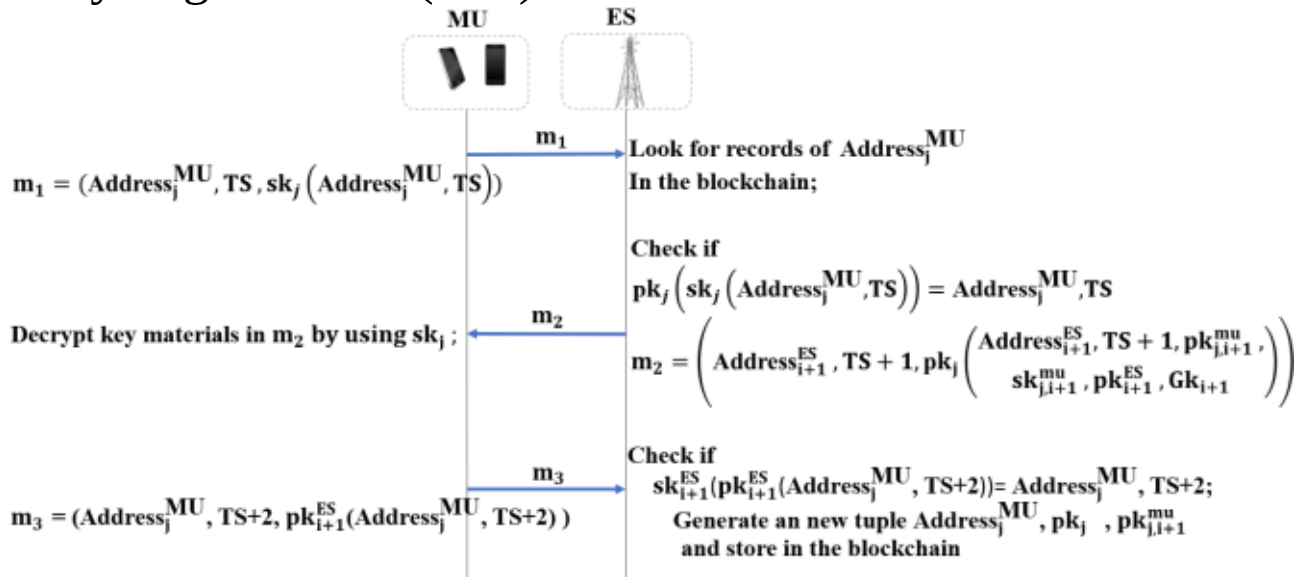

FIGURE 4. **Authentication Process Between MU and ES During Server Change**

The block validation process is divided into two stages:

**First Validation Stage:** When an edge server (ES) from group $i$ generates a new block, it must first be validated within group $i$ before it can be broadcast to the entire network. Each member of the group verifies the transactions included in the block. After the block is validated by the mobile users (MUs), they sign the block's hash using their private keys, $sk_{j,i}^{mu}$. The block can only advance to the second validation stage, involving other groups, if it has received signatures from all members of group $i$. To streamline the process, the Boneh–Lynn–Shacham (BLS) signature algorithm is utilized to consolidate multiple signatures into a single one. Furthermore, the group public key GK$i$ is computed by aggregating the public keys of all members within group $i$, as detailed in (1) [32].

$$Gk_i = \mathrm{f}(pk_{1,i}^{mu}, pk_{2,i}^{mu}, pk_{3,i}^{mu}, \dots, pk_{j,i}^{mu}) \tag{1}$$

Since mobile nodes are continually moving, a mobile node may join the group after authenticating with the server, or some existing members may leave. In such cases, the edge server (ES) in the group must re-collect the public keys from each member to update the group public key GK$i$. Given the ES's ample computational resources, the time required for this re-aggregation is minimal and does not impact the update of the group public key or the subsequent consensus process. Once the latest proposed block from this group is confirmed and added to the blockchain by other groups, the ES will update the group public key $GK_i$ on the blockchain if there are changes in the group composition.

**Second Validation Stage:** When a proposed block from group $i$ passes the first validation stage, it is broadcast to other groups for the second validation stage. Upon receiving the block, mobile users (MUs) in other groups do not need to re-verify all the transactions since they have already been validated. Instead, they focus on verifying the aggregated signature on the block's hash. During this process, the MUs retrieve GK$i$, the most recent group public key for group $i$ stored on the blockchain, and use it to validate the aggregated signature. Once the second validation stage is completed, the proposed block from group $i$ is deemed valid and is added to the blockchain.

### C. Content Model

In this study, we adopt the content model outlined in [10]. The set of contents requested by mobile users (MUs) is represented as $M_i = \{m_{i1}, m_{i2}, m_{i3}, \dots, m_{ij}, \dots, m_{iN_i(t)}\}$, where $N_i(t)$ denotes the number of MUs connected to edge server $i$ during time period $t$, and $m_{ij}$ is the content requested by MU$j$ from edge server $i$. To enable more efficient content caching, the content $m_{ij}$ is encoded using a layered coding approach such as Scalable Video Coding (SVC), which allows partial caching of the content on edge servers. The total size of the content $m_{ij}$ is denoted by $S_{ij}$, with the calculation formula provided in (2) [10].

$$S_{ij} = \sum_{k=1}^{K_{ij}} s_{ijk}, \forall i \in \mathrm{I}, \forall j \in \mathrm{J} \tag{2}$$

where $s_{ijk}$ represents the size of the $k$-th layer of content $m_{ij}$, and $k_{ij}$ is the total number of layers of content $m_{ij}$. In the proposed method, each content layer is delivered sequentially from the edge server (ES) to the mobile users (MUs).

Content popularity adheres to a Zipf distribution. The popularity of content $m_{ij}$, denoted as $f_{ij}$, is ranked in descending order of access frequency over a given period, as calculated by (3) [10]:

$$f_{ij} = \frac{1}{(\tau(m_{ij}))^{\gamma}} \cdot \frac{1}{\sum_{j=1}^{N_i(t)} (\frac{1}{j})^{\gamma}}, \forall i \in \mathrm{I}, m_{ij} \in M_i \tag{3}$$

where $\gamma$ is a positive parameter that influences the skewness of the request distribution. A larger $\gamma$ means that popular content makes up a significant portion of the requests. Furthermore, content popularity reflects the likelihood of a content request.

### D. Cache Token Creation

To safeguard against man-in-the-middle attacks and data manipulation during cache requests from $MU_j$ to $ES_i$, a cache token is generated. The method for creating the cache token follows the approach outlined in [10]. Here's how it works: Before requesting cache resources from $ES_i$, $MU_j$ generates a cache token, $\delta_{ij}$, which comprises $\delta_{ij} = (ID(m_{ij}), k_{ij}^*, h_{ij})$. In this token ID ($m_{ij}$) is the identifier for the content $m_{ij}$, $k_{ij}^*$ represents the total number of content layers cached by $MU_j$ at $ES_i$ and $h_{ij} = H(ID(m_{ij}))$ is the hash value produced by the public hash function $H$.

$MU_j$ then signs $\delta_{ij}$ with its private key, denoted as $sign(\delta_{ij})$. $MU_j$ sends both $sign(\delta_{ij})$ and $\delta_{ij}$ along with a timestamp $TS$ to $ES_i$, while retaining $h_{ij}$. Upon receiving $\delta_{ij}$, $ES_i$ verifies the authenticity of the data by checking the signature with $MU_j$'s public key. This verification involves validating sign ($\delta_{ij}$). If the signature is verified successfully, $ES_i$ then confirms $h_{ij}$ by comparing hash values. Once these checks are complete, $\delta_{ij}$ is stored in the cache memory. Following these steps, $ES_i$ records the cache transaction by (4). The edge server then forwards this transaction to the associated smart contract, and the transaction is recorded on the blockchain.

$$trans_{cache} = (h_{ij}, sign(\delta_{ij}), \delta_{ij}, Address_i^{ES}) \tag{4}$$

### E. Edge Server's Price Model

Given the limited cache capacity of edge servers, MUs must pay a fee to access the caching services. To determine the cost of using these services, this study utilizes a pricing mechanism similar to the one described in [10], with some adjustments to the price parameters. The calculation formula is detailed in (5) [10].

$$p_i^* = \begin{cases} \dfrac{C_i + c_i\xi_i}{2\xi_i}, C_i \geq c_i\xi_i \\ c_i, C_i < c_i\xi_i \end{cases} \tag{5}$$

where $C_i$ represents the cache capacity of the edge server, $c_i$ is the cost per unit of cached data, and $\xi_i$ is the pricing parameter of the edge server. The parameter $\xi_i$ is influenced by factors such as the cache demand from each MU, the size of the requested content, the number of MUs, and the content's popularity. The calculation of $\xi_i$ is given in (6).

$$\xi_i = log(1 + \frac{\rho_i \sum_{\hat{\tau}=1}^{t-1} e^{\eta(\hat{\tau}-t)}}{E(d_i)E(S_i)\sum_{\hat{\tau}=1}^{t-1} N_i(\hat{\tau})e^{\eta(\hat{\tau}-t)}}) \qquad (6)$$

where $\rho_i$ is the adjustment parameter, $\eta$ is the attenuation parameter, $N_i(\hat{\tau})$ denotes the number of MUs requesting cache resources from edge server $i$ within the time slot $\hat{\tau}$, $E(d_i)$ represents the cache demand from each MU for edge server $i$, and $E(S_i)$ is the average size of content requested from all contents stored on edge server $i$.

The cache demand of each Mobile User (MU) significantly impacts the overall cache demand on Edge Servers (ESs). An increase in the cache demand from individual MUs leads to a corresponding rise in the total demand on the ESs. Since ESs and MUs operate independently, the cache demand of each MU is considered private and cannot be directly observed by the ES. Consequently, each ES must rely on historical demand data from MUs to predict and estimate current cache demands. It is assumed that the cache demand for each MU is uniformly distributed within the interval $[\underline{d}_i, \bar{d}_i]$ [10]. The cache demand for each MU is calculated using (7) [10].

$$E(d_i) = \frac{(\underline{d}_i + \bar{d}_i)}{2} \qquad (7)$$

When MUs request content, larger content sizes typically lead them to cache more layers on the edge server to minimize delivery latency [10]. The average content size for a demand across all contents on Edge Server (ES) $i$ is calculated using (8) [10].

$$E(S_i) = \sum_{j=1}^{N_i} f_{ij} s_{ij} \qquad (8)$$

### F. Edge Server Cache Resource Allocation

In the proposed method, each mobile user (MU) communicates their cache demand to the associated edge server (ES). Given the variability in cache demands and the limited capacity of the server, the Max-min fairness mechanism [10] is employed to ensure fair allocation of cache resources among MUs. This approach guarantees that every MU receives a specific portion of cache resources.

If the total cache demand from MUs is within the ES's cache capacity, each MU receives the full amount of cache resources they requested. Conversely, if the total demand surpasses the ES's capacity, the cache resources are distributed such that each MU gets a proportional share [10].

As illustrated in Figure 5, each mobile user is allocated a defined amount of cache resources. In this algorithm, $K_i^* = \{k_{i1}^*, k_{i2}^*, \ldots, k_{iN_i(t)}^*\}$ denotes the set of content layers requested by MUs connected to ES $i$. Meanwhile, $K_i^{**} = \{k_{i1}^{**}, k_{i2}^{**}, \ldots, k_{iN_i(t)}^{**}\}$ represents the set of content layers that ES $i$ allocates to the connected MUs.

1: **Input:** $C_i$, $K_i^* = \{k_{i1}^*, k_{i2}^*, \ldots, k_{iN_i(t)}^*\}$
2: **Output:** $K_i^{**} = \{k_{i1}^{**}, k_{i2}^{**}, \ldots, k_{iN_i(t)}^{**}\}$
3: **Initialize:** $\forall j$, $k_{ij} = 1$ or $k_{ij} = 0$ , $k = C_i / N_i(t)$
4: **while** $min\{d_{ij}, j \in N_i\} < k$ and $\exists k_{ij} > 0$ $d_{ij} = s_{ijk}$ **do**
5: **repeat**
6: **for** each $d_{ij}$ **do**
7: **if** $d_{ij} \leq k$ **then**
8: $\Xi_{sat} \longleftarrow d_{ij}$, where $\Xi_{sat}$ is the set of demands that have been satisfied.
9: **end if**
10: **end for**
11: **if** $\Xi_{sat} \neq \emptyset$ **then** $C_i = C_i - \sum_{d_{ij} \epsilon \Xi_{sat}} d_{ij}$ , $D_i = D_i / \Xi_{sat}$ ,
12: $N_i(t) = N_i(t) - |\Xi_{sat}|$, $\Gamma_{sat} = \Gamma_{sat}\ U\ \Xi_{sat}$ , $\Xi_{sat} \neq \emptyset$ , where $\Gamma_{sat}$ is other temporary set of demands that have been satisfied
13: **if** $N_i \neq 0$ **then**
14: $k = C_i / N_i$
15: **end if**
16: **else**
17: $N_i(t) = N_i(t) - 1$, $k = C_i / N_i(t)$.
18: **end if**
19: **until** $N_i = 0$.
20: **for** $d_{ij}$ **do**
21: **if** $d_{ij} \in \Gamma_{sat}$ and $k_{ij} \neq k_{ij}^*$ **then**
22: $k_{ij} = k_{ij} + 1$, $d_{ij} = d_{ijk_{ij}}$ .
23: **else if** $d_{ij} \in \Gamma_{sat}$ and $k_{ij} = k_{ij}^*$ **then**
24: $k_{ij}^{**} = k_{ij}$ , $k_{ij} = 0$.
25: **end if**
26: **end for**
27: $\Gamma_{sat} \neq \emptyset$, $\Xi_{sat} \neq \emptyset$, $k = C_i / N_i(t)$.
28: **end while**
29: $k_{ij}^{**} = k_{ij}$.

**FIGURE 5. Max-min based Fair Caching Resource Allocation Algorithm [10]**

The algorithm in Figure 5 describes a Max-Min Fairness-based caching resource allocation mechanism. The goal is to ensure that mobile users receive a fair share of cache resources when the edge server (ES) capacity is limited. The algorithm begins by allocating the minimum required layers to each mobile user and iteratively increases allocations while ensuring that no user is starved. The algorithm stops when either the cache is fully allocated or no further fair allocation is possible.

Let $N$ denote the number of mobile users connected to a specific ES, and $L$ denote the total number of content layers requested. In the worst case, the algorithm must iterate through all users and requested layers. Thus, the time complexity is $O(N \cdot L)$, which is acceptable for mobile edge environments where both N and L are relatively small due to local constraints. The algorithm is lightweight and scalable, making it suitable for real-time cache allocation in edge servers.

### G. Trust Model

Throughout this work, $t$ denotes intermediate trust evaluations obtained either directly from user interactions or indirectly via third-party observations. In contrast, $T$ refers to the aggregated final trust score after applying credibility and weighting mechanisms. This distinction ensures that raw feedback is properly processed before influencing trust-based decisions.

#### 1) IMPACT OF THE PARAMETER F IN TRUST COMPUTATION

The parameter f functions as a trust decay coefficient, determining how recent interactions influence the final trust score. A higher $f$ (e.g., 0.9) prioritizes recent feedback, enabling rapid adaptation to behavioral changes. Conversely, a lower $f$ (e.g., 0.5) integrates longer-term historical data, providing stability. In dynamic environments with frequent node behavior variation, higher $f$ values enhance responsiveness, while lower values improve resistance to temporary noise.

In the network, some edge servers (ESs) may engage in malicious activities, undermining the delivery of secure caching services to mobile users (MUs) through various attacks. To mitigate performance degradation and maintain service quality, we employ trust management to ensure reliable caching services [10]. This paper uses trust as a metric to assess the reliability and credibility of an ES in providing secure caching services.

To calculate trust, we utilize the mechanism described in [23], with minor adjustments in the calculation of indirect trust. This mechanism operates in two stages: 1) Direct Trust Calculation: involves evaluating the experiences of MUs who interact directly with the ES. 2) Indirect Trust Calculation: considers the evaluations of mobile users regarding edge servers they have interacted with previously.

#### 2) USER EVALUATION CREDIBILITY

To analyze the resource cost of the two-stage consensus protocol, we measured the average CPU usage (3–5%), memory overhead (<20MB), and message latency (avg. 2.1s) during simulation. Unlike conventional PoW/PoS schemes, our system avoids energy-intensive operations on mobile nodes. Consensus is limited to edge servers using digital signature verification and lightweight hash comparisons. This design offers practical deployability on real-time edge infrastructures.

To justify the selected parameters, we evaluated model performance under varying initial trust thresholds $\rho \in \{0.05, 0.1, 0.2\}$ and indirect trust weights $\alpha \in \{0.7, 0.8, 0.9\}$. Results indicate that $\rho = 0.1$ and $\alpha = 0.9$ deliver the best trade-off between false positives and detection latency. The model maintains over 89% accuracy across all tested combinations, confirming the robustness of the approach to parameter variation.

The incentive mechanism promotes sharing behavior by rewarding nodes based on trust-adjusted contributions. Simulations under cooperative and competitive modes show that allowing nodes to rate content quality enhances trust calibration. This results in a 12–18% improvement in cache hit ratio and throughput compared to scenarios without feedback-based incentives. Thus, incorporating feedback into the incentive loop directly improves cooperation and content dissemination efficiency.

To evaluate the system's robustness, we tested trust evaluation under different levels of user feedback noise: 0%, 10%, 20%, and 30%. Our model achieved 94.2%, 91.5%, 89.3%, and 87.1% accuracy, respectively. Despite increased uncertainty, the layered trust mechanism and credibility filtering preserved reliable detection. This validates the system's resilience in dynamic, noisy environments.

#### 3) ENERGY CONSUMPTION ANALYSIS

To evaluate the energy cost, we estimate the power used for trust evaluation and communication. Each trust evaluation requires lightweight arithmetic operations (~10 µJ), while message exchange for consensus involves small packets over short-range wireless (~1–2 mJ per transaction). Given the infrequent updates and localized processing, the overall energy overhead remains under 5% of typical mobile device battery capacity for moderate participation rates. This confirms the model's suitability for real-world deployment without excessive energy drain.

Malicious MUs may collude to lower the trustworthiness of reliable ESs or inflate the trust of malicious ESs. To counteract such actions, the credibility of MU evaluations is assessed [23]. Equation (9) details the method for determining the credibility of MU evaluations [23].

$$RE_{ji} = e^{(-SD_{ji})} \tag{9}$$

where $SD_{ji}$ represents the standard deviation of the credibility of MUs' evaluations, as used in (10) [23].

$$SD_{ji} = \sqrt{\frac{\sum_{s \in N(p,q)} (E_{js} - \bar{N})^2}{N(p,q)}} \tag{10}$$

where $E_{js}$ $(0 \le E_{js} \le 1)$ denotes $MU_j$'s evaluation of the services provided by ES s, $N(p, q)$ is the set of all ESs evaluated by $MU_j$, and $\bar{N}$ is the average of $N(p, q)$, as used in (11) [23].

$$\bar{N} = \frac{1}{c} \sum N(p,q) \tag{11}$$

where $c$ represents the length of the set $N(p, q)$.

#### 4) DIRECT TRUST

Malicious edge servers (ESs) may initially appear highly reliable by delivering excellent performance and quickly gaining a high level of trust from minimal interactions. However, once they have established this trust, their service quality often deteriorates significantly, posing serious security risks and potentially compromising overall network security. To counteract this, the correction parameter $f$ is introduced to limit the trust accumulation of ESs that have had only a few successful interactions. This ensures that ESs can only build higher trust ratings by consistently providing high-quality services. The total number of successful interactions between ES$i$ and mobile user $j$ is denoted by SIC. The parameter $f$ is determined as follows [23]:

$$f = \sqrt{\frac{SIC}{SIC+1}} \tag{12}$$

Direct trust is computed using (13):

$$DT_{ji} = f \times RE_{ji} \times E_{ji}, 0 \le DT_{ji} \le 1 \tag{13}$$

#### 5) INDIRECT TRUST

This evaluates the assessments of mobile users regarding the edge servers with which they have previously interacted. The formula for calculating indirect trust is as follows:

$$NT_i = \frac{1}{n} \times \sum_{j=1}^{n} \sum_{\hat{\tau}=1}^{t-1} DT_{ji}(\hat{\tau})\, e^{\eta(\hat{\tau}-t)}, 0 \le NT_i \le 1 \tag{14}$$

where $n$ is the total number of MUs who have previously interacted with $ES_i$, and $\tau$ is the time period used to assess the recency of each MU's evaluation of $ES_i$.

#### 6) EDGE SERVER TRUST

This reflects the satisfaction level of $MU_j$ with the services provided by $ES_i$. The total trust in $ES_i$ is calculated using (15) [23].

$$T_i = \alpha \times DT_{ji} + (1-\alpha) \times NT_i, 0 \le T_i \le 1 \tag{15}$$

where $\alpha$ represents the weighting parameter.

To reduce the impact of potentially manipulated user feedback, the system calculates a credibility score for each mobile user's evaluation using Equation (9). This score reflects the consistency of a user's feedback across multiple edge servers. In addition, Equation (15) assigns a higher weight to indirect trust ($\alpha = 0.9$), which aggregates evaluations from multiple sources, thereby mitigating the influence of any single biased or malicious user. This layered approach enhances robustness against collusion and bad-mouthing attacks.

After obtaining the trust level of each edge server, the mobile node selects the optimal edge server from the available servers in its vicinity for receiving the desired cache service based on two criteria: 1) The edge server must be trustworthy. 2) The cache price of the edge server should be low, and its capacity should be high [10]. Thus, Each MU initially sets a trust threshold to assess the trustworthiness of the ES, determined by (16).

$$\varrho(t) = \beta \bar{T}(t-1) + (1-\beta)\varrho(t-1) \tag{16}$$

where $\varrho$ is the trust threshold, $\bar{T}$ is the average trust level of all available ESs, and $\beta$ is the weighting parameter. The average trust $\bar{T}$ is calculated using (17).

$$\bar{T}(t-1) = \frac{1}{m}\sum_{i=1}^{m} T_i(t-1) \tag{17}$$

Each MU then evaluates the cache status of the ESs to determine the optimal one. The trust, price, and remaining cache capacity of ES i are retrieved from the blockchain using the public key of this server. For $MU_j$, the cache status of $ES_i$ is calculated using (18) [10].

$$\Delta_i = \frac{\omega T_i(t-1) + (1-\omega)C_i}{p_i^*} \tag{18}$$

where $\omega$ is the weighting parameter. The MU then selects the optimal ES from the available servers in its vicinity. The optimal ES is identified using (19).

$$i^* = \underset{i}{argmax}\{\Delta_i | T_i(t-1) \ge \varrho(t)\} \tag{19}$$

Following the provision of service to the mobile node, the trust level, cost, and remaining cache capacity of the optimal ES are calculated and subsequently stored on the blockchain. The trust levels of ESs are periodically updated on the blockchain. ESs with trust scores below the threshold are excluded from selection and effectively removed. This threshold is also regularly reassessed and updated.

### *H. Cooperative Mobile Node Reward Model*

When an ES is either unwilling or unable to share its cache capacity, it seeks to access the cache resources of nearby MUs. Given that MUs are generally reluctant to share their cache due to their limited capacity and selfish tendencies, this study introduces a reward mechanism to incentivize sharing.

In this approach, each mobile node declares the amount of cache capacity it is willing to share. Nodes that agree to share their cache are termed as cooperative mobile nodes. These caching mobile users (CMUs) are then rewarded based on their trust level and the amount of cache they provide.

The rewards are distributed as tokens, which can be accumulated by the cooperative mobile nodes. These tokens can be used to either reduce service costs from ESs or potentially receive services for free.

#### 1) SELECTION OF COOPERATIVE MOBILE NODES

Initially, the ES establishes a trust threshold using Equation (16) to determine the trustworthiness of potential cooperative mobile nodes. After setting this threshold, the ES evaluates the cache status of each cooperative mobile node to identify the optimal candidates. The cache status for a cooperative mobile node $h$ on $ES_i$ is computed using (20).

$$\Delta_h = \tilde{\alpha} c_h + (1-\tilde{\alpha}) T_h(t-1) \tag{20}$$

where $c_h$ represents the amount of capacity shared by the cooperative mobile node, $T_h$ denotes the trust level of the cooperative mobile node, and $\tilde{\alpha}$ is the weighting parameter. The optimal cooperative mobile node selected by $ES_i$ is determined using (21).

$$h^* = \underset{h}{argmax}\{\Delta_h | T_h(t-1) \geq \varrho(t)\} \quad (21)$$

The trust levels of cooperative mobile nodes are updated periodically on the blockchain. Nodes with trust levels below the threshold are excluded from the selection process and removed. The trust threshold for these nodes is also regularly reviewed and adjusted.

#### 2) ALLOCATION OF REWARDS TO COOPERATIVE MOBILE NODES

The trust of a cooperative node is assessed using Equation (15). The rewards allocated to the cooperative mobile node are calculated according to (22).

$$Re_h = r \times c_h \times T_h(t) \quad (22)$$

where $r$ represents the reward given per unit of data. In the proposed method, each unit of reward is exchanged for one token. Cooperative mobile nodes can use these tokens to lower the cost of services from edge servers.

```
1: Input: p_i
2: Output: d_ij*
3: Initialize: t̂ = 0
4: repeat
5:   if p_i ≥ (w_ij τθ/d_ij(4 + 2θ)) then
6:     d_ij* = 0.
7:   else
8:     Each mobile user updates its demand by (23).
9:     t̂ = t̂ + 1.
10:  end if
11: until d_ij keeps unchanged, and d_ij* = d_ij(t̂).
```

FIGURE 6. **Gradient Descent based Optimal Caching Demand Seeking Algorithm**

### *I. Optimal Cache Demand of Mobile Users from Edge Servers*

The approach outlined in [10] is adapted with slight modifications to calculate each MU's optimal cache demand from the ES. The cache demand of MUs is directly affected by the ES's pricing. When cache prices are high, MUs will request less cache to reduce costs. Conversely, when prices are lower, users are more likely to request additional layers of content to improve their overall experience [10]. The optimal cache demand for each mobile user is determined using (23) and the algorithm presented in Figure 6.

$$d_{ij}(\hat{t}+1) = d_{ij}(\hat{t}) + \upsilon\left(\frac{\partial u_{ij}(\hat{t})}{\partial d_{ij}(\hat{t})}\right) = d_{ij}(\hat{t}) - p_i\upsilon \quad (23)$$

where $\upsilon$ denotes the iteration rate of cache demand, $d_{ij}$ represents the cache demand of MU$j$ from ES$i$ at stage $\hat{t}$, and $u_{ij}$ signifies the utility of MU$_j$ connecting to ES$_i$. The utility is calculated using (24).

$$u_{ij} = w_{ij}\, log\left(1 + \frac{\theta}{1+e^{-\tau}}\right) - p_i d_{ij}, \forall i \in \mathrm{I}, \forall j \in N_i \quad (24)$$

where $\theta$ and $\tau$ are metrics used to quantify the Quality of Experience (QoE). The variable $w_{ij}$ represents the willingness of MU$_j$ to cache content on ES$_i$, which is related to the content's popularity. Higher content popularity, denoted as $m_{ij}$, leads to the MU caching more layers of the content on the ES. Equation (25) provides the formula for calculating $w_{ij}$ [10].

$$w_{ij} = \frac{\hat{\alpha}}{1+e^{-\hat{\beta} f_{ij}}} \quad (25)$$

where $\hat{\alpha}$ and $\hat{\beta}$ are predetermined constants selected based on user preferences. After calculating the optimal cache demand, each MU can send their demand to the corresponding ES.

### *J. Design Justification and Operational Characteristics*

This section presents a structured empirical analysis of the core design aspects of the proposed system, including resource efficiency, trust parameter calibration, energy consumption, incentive mechanisms, and the model's resilience to noisy user feedback.

- **Resource Utilization Analysis of the Two-Stage Consensus Protocol:** To analyze the resource cost of the two-stage consensus protocol, we measured the average CPU usage (3–5%), memory overhead (<20MB), and message latency (avg. 2.1s) during simulation. Unlike conventional PoW/PoS schemes, our system avoids energy-intensive operations on mobile nodes. Consensus is limited to edge servers using digital signature verification and lightweight hash comparisons. This design offers practical deployability on real-time edge infrastructures.
- **Empirical Evaluation and Justification of Trust Parameter Configuration:** To justify the selected parameters, we evaluated model performance under varying initial trust thresholds $\varrho \in \{0.05, 0.1, 0.2\}$ and indirect trust weights $\alpha \in \{0.7, 0.8, 0.9\}$. Results indicate that $\varrho = 0.1$ and $\alpha = 0.9$ deliver the best trade-off between false positives and detection latency. The model maintains over 89% accuracy across all tested combinations, confirming the robustness of the approach to parameter variation.
- **Impact of Incentive Mechanism and Feedback:** The incentive mechanism promotes sharing behavior by rewarding nodes based on trust-adjusted contributions. Simulations under cooperative and competitive modes show that allowing nodes to rate content quality enhances trust calibration. This results in a 12–18% improvement in cache hit ratio and throughput compared to scenarios without feedback-based incentives. Thus, incorporating feedback into the incentive loop directly improves cooperation and content dissemination efficiency.
- **Trust Model Stability Under Noisy Feedback Conditions:** To evaluate the system's robustness, we tested trust evaluation under different levels of user feedback noise: 0%, 10%, 20%, and 30%. Our model achieved 94.2%, 91.5%, 89.3%, and 87.1% accuracy, respectively. Despite increased uncertainty, the layered trust mechanism and credibility filtering preserved reliable detection. This validates the system's resilience in dynamic, noisy environments.

- **Energy Consumption Analysis:** To evaluate the energy cost, we estimate the power used for trust evaluation and communication. Each trust evaluation requires lightweight arithmetic operations (~10 μJ), while message exchange for consensus involves small packets over short-range wireless (~1–2 mJ per transaction). Given the infrequent updates and localized processing, the overall energy overhead remains under 5% of typical mobile device battery capacity for moderate participation rates. This confirms the model's suitability for real-world deployment without excessive energy drain.
- **Impact of the Parameter *f* in Trust Computation:** The parameter *f* functions as a trust decay coefficient, determining how recent interactions influence the final trust score. A higher *f* (e.g., 0.9) prioritizes recent feedback, enabling rapid adaptation to behavioral changes. Conversely, a lower *f* (e.g., 0.5) integrates longer-term historical data, providing stability. In dynamic environments with frequent node behavior variation, higher *f* values enhance responsiveness, while lower values improve resistance to temporary noise.

For ease of reference, the key notations employed in this paper are summarized in Table 2.

TABLE 2
SUMMARY OF SYMBOLS

| Symbol | Description |
|---|---|
| $Address_j^{MU}$ | Address of mobile user *j* |
| $Address_i^{ES}$ | Address of edge server *i* |
| $pk_j$ | Public key of mobile user *j* |
| $sk_j$ | Private key of mobile user *j* |
| $pk_{AS}$ | Public key of Authentication Server |
| $sk_{AS}$ | Private key of Authentication Server |
| $GK_i$ | Group public key of group *i* |
| $pk_{j,i}^{mu}$ | Public key of mobile user *j* in group *i* |
| $sk_{j,i}^{mu}$ | Private key of mobile user *j* in group *i* |
| $pk_i^{ES}$ | Public key of edge server *i* |
| $sk_i^{ES}$ | Private key of edge server *i* |
| $N_i(t)$ | Number of mobile users connected to edge server *i* during time *t* |
| $m_{ij}$ | Content requested by mobile user *j* from edge server *i* |
| $S_{ij}$ | Total size of content $m_{ij}$ |
| $K_{ij}$ | Total number of layers of content $m_{ij}$ |
| $s_{ijk}$ | Size of the k-th layer of content $m_{ij}$ |
| $f_{ij}$ | Popularity of content $m_{ij}$ |
| $p_i^*$ | Price of edge server *i* |
| $C_i$ | Cache capacity of edge server *i* |
| $c_i$ | Cost of caching per unit of data on edge server *i* |
| $\xi_i$ | Pricing parameter of edge server *i* |
| $E(d_i)$ | Cache demand of mobile user *j* from edge server *i* |
| $E(s_i)$ | Average size of content for a request on edge server *i* |
| $E_{ji}$ | Evaluation of user *j* regarding services provided by edge server *i* |
| $RE_{ji}$ | Credibility of evaluations made by mobile user |
| $DT_{ji}$ | Direct trust of edge server *i* |
| $NT_i$ | Indirect trust of edge server *i* |
| $T_i$ | Total trust of edge server *i* |
| $\varrho(t)$ | Trust threshold |
| $\Delta_i$ | Cache status of edge server *i* |
| $Re_h$ | Reward granted to cooperative mobile node *h* |
| $d_{ij}$ | Optimal cache demand of mobile user *j* from edge server *i* |
| $u_{ij}$ | Utility of mobile user *j* connecting to edge server *i* |
| $w_{ij}$ | Willingness of mobile user *j* to cache content on edge server *i* |

## V. EVALUATION

This section presents the simulation results evaluating the performance of the proposed approach. The main objectives include ensuring security, accurate classification of edge servers, and robust trust management to withstand malicious attacks and the specific challenges of edge computing environments. To achieve comprehensive insights, well-established metrics such as accuracy, recall, specificity, and F1-score were employed.

### *A. Simulation Setup*

The proposed approach was implemented through the utilization of the Python programming language, Solidity, and the Ganache test network.

To assess the effectiveness of the proposed model in identifying trustworthy and malicious edge servers, we used standard classification metrics: accuracy, precision, recall, specificity, and F1-score [33]. These metrics were computed using a three-class confusion matrix, which distinguishes between high-quality (HQ), low-quality (LQ), and malicious (M) edge servers.

To ensure fair evaluation across all classes, we applied macro-averaging, which gives equal weight to each class regardless of its sample size. This approach allows for a balanced performance analysis even in the presence of class imbalance.

#### 1) DETERMINING WEIGHT PARAMETERS

This section aims to determine the optimal weight parameters for the proposed method using a confusion matrix. The simulation parameters were adjusted according to Table 3 to identify these optimal weights. In each simulation time, a randomly selected mobile user (MU) connects to the most optimal edge server (ES) available in its vicinity to receive services. For all evaluations of the proposed approach, it is assumed that the number of malicious servers remains below 50% of the total simulated servers.

TABLE 3
SIMULATION PARAMETERS USED TO DETERMINE THE OPTIMAL WEIGHT.

| Parameter | Value |
|---|---|
| Number of Edge Servers | 10 |
| Number of Mobile Users | 20 |
| Malicious Edge Servers | 20% |
| Low-Quality Edge Servers | 20% |
| High-Quality Edge Servers | 60% |
| Initial Edge Server Trust Value | 0.8 |
| Initial Edge Server Trust Threshold | 0.1 |
| Simulation Time | 500 interactions |
| $\hat{\tau}$ | 100 seconds |
| $\rho_i$ | $10^6$ |
| $c_i$ | 100 |
| Number of Runs | 10 |

The trust that a mobile user (MU) assigns to an edge server (ES) is calculated using (15). This section examines three candidate ranges, as outlined in Table 4, to determine the trust value assigned by the MU to the ES.

TABLE 4
CANDIDATE RANGES FOR DETERMINING TRUST IN THREE CLASSES HQ. LQ AND M.

| 1 | 2 | 3 |
|---|---|---|
| $T = \begin{cases} M & if\ \ T < 0.2 \\ LQ & if\ 0.2\ \leq T < 0.4 \\ HQ & if\ 0.4 \leq T < 1 \end{cases}$ | $T = \begin{cases} M & if\ \ T < 0.3 \\ LQ & if\ 0.3\ \leq T < 0.5 \\ HQ & if\ 0.5 \leq T < 1 \end{cases}$ | $T = \begin{cases} M & if\ \ T < 0.3 \\ LQ & if\ 0.3\ \leq T < 0.6 \\ HQ & if\ 0.6 \leq T < 1 \end{cases}$ |

Tables 5, 6, and 7 illustrate the impact of increasing the trust weight on the confusion matrix metrics for three server statuses: malicious, low-quality, and high-quality across the three specified ranges. Since these metrics are typically used for two-class classification, and the proposed approach employs a three-class classification, a macro-averaging method was used to calculate the confusion matrix metrics using the values from Tables 5, 6, and 7. Figures 7, 8, and 9 depict the macro-averaged precision, recall, specificity, and F1-score from Tables IV, V, and VI, respectively. By applying these metrics across distinct server classes, the model can robustly handle security threats such as content tampering and collusion, which are prevalent in edge caching.

TABLE 5
CONFUSION MATRIX METRICS VALUES FOR RANGE 1

| $\alpha$ | Accuracy | Precision | | | Recall | | | Specificity | | | F1-score | | |
|---|---|---|---|---|---|---|---|---|---|---|---|---|---|
| | | M | LQ | HQ | M | LQ | HQ | M | LQ | HQ | M | LQ | HQ |
| 0.1 | 0.8375 | 0.875 | 0.6333 | 1 | 1 | 0.8125 | 0.7916 | 0.9531 | 0.8437 | 1 | 0.9249 | 0.6714 | 0.8742 |
| 0.2 | 0.7625 | 0.7712 | 0.4562 | 1 | 1 | 0.625 | 0.7291 | 0.9062 | 0.7968 | 1 | 0.8583 | 0.5126 | 0.8329 |
| 0.3 | 0.7875 | 0.8125 | 0.4416 | 1 | 1 | 0.6875 | 0.7504 | 0.9218 | 0.8125 | 1 | 0.8833 | 0.5277 | 0.8420 |
| 0.4 | 0.775 | 0.875 | 0.5275 | 1 | 1 | 0.8125 | 0.6870 | 0.9531 | 0.7656 | 1 | 0.9249 | 0.6011 | 0.8023 |
| 0.5 | 0.7125 | 0.7295 | 0.3895 | 1 | 1 | 0.5625 | 0.6662 | 0.8906 | 0.75 | 1 | 0.8333 | 0.4482 | 0.7838 |
| 0.6 | 0.7625 | 0.8754 | 0.4562 | 0.975 | 1 | 0.75 | 0.6870 | 0.9531 | 0.7656 | 0.9687 | 0.9249 | 0.5547 | 0.7985 |
| 0.7 | 0.7375 | 0.8333 | 0.3625 | 0.975 | 1 | 0.625 | 0.6870 | 0.9218 | 0.7656 | 0.9687 | 0.8920 | 0.4547 | 0.7985 |
| 0.8 | 0.6625 | 0.7374 | 0.26 | 1 | 0.9375 | 0.5625 | 0.6037 | 0.875 | 0.7031 | 1 | 0.7880 | 0.3541 | 0.7344 |
| 0.9 | 0.625 | 0.5833 | 0.2083 | 0.9375 | 1 | 0.3125 | 0.6045 | 0.8125 | 0.7343 | 0.9687 | 0.7337 | 0.25 | 0.7282 |

Macro-averaging is used in this evaluation to ensure that all categories—malicious, low-quality, and high-quality—are equally represented, which provides an unbiased view of the model's classification effectiveness across varying server types.

TABLE 6
CONFUSION MATRIX METRICS VALUES FOR RANGE 2

| $\alpha$ | Accuracy | Precision | | | Recall | | | Specificity | | | F1-score | | |
|---|---|---|---|---|---|---|---|---|---|---|---|---|---|
| | | M | LQ | HQ | M | LQ | HQ | M | LQ | HQ | M | LQ | HQ |
| 0.1 | 0.6 | 0.7592 | 0.2740 | 1 | 1 | 0.1 | 0.6 | 0.7592 | 0.2740 | 1 | 1 | 0.1 | 0.6 |
| 0.2 | 0.5666 | 0.7407 | 0.2444 | 1 | 1 | 0.2 | 0.5666 | 0.7407 | 0.2444 | 1 | 1 | 0.2 | 0.5666 |
| 0.3 | 0.6222 | 0.6481 | 0.2055 | 1 | 1 | 0.3 | 0.6222 | 0.6481 | 0.2055 | 1 | 1 | 0.3 | 0.6222 |
| 0.4 | 0.5333 | 0.6296 | 0.1814 | 1 | 1 | 0.4 | 0.5333 | 0.6296 | 0.1814 | 1 | 1 | 0.4 | 0.5333 |
| 0.5 | 0.5888 | 0.6555 | 0.1944 | 1 | 1 | 0.5 | 0.5888 | 0.6555 | 0.1944 | 1 | 1 | 0.5 | 0.5888 |
| 0.6 | 0.4444 | 0.6481 | 0.1280 | 0.7407 | 1 | 0.6 | 0.4444 | 0.6481 | 0.1280 | 0.7407 | 1 | 0.6 | 0.4444 |
| 0.7 | 0.5444 | 0.5555 | 0.1851 | 1 | 1 | 0.7 | 0.5444 | 0.5555 | 0.1851 | 1 | 1 | 0.7 | 0.5444 |
| 0.8 | 0.5888 | 0.6111 | 0.2722 | 0.9777 | 0.8888 | 0.8 | 0.5888 | 0.6111 | 0.2722 | 0.9777 | 0.8888 | 0.8 | 0.5888 |
| 0.9 | 0.5777 | 0.5925 | 0.1740 | 0.9777 | 1 | 0.9 | 0.5777 | 0.5925 | 0.1740 | 0.9777 | 1 | 0.9 | 0.5777 |

TABLE 7
CONFUSION MATRIX METRICS VALUES FOR RANGE 3

| $\alpha$ | Accuracy | Precision | | | Recall | | | Specificity | | | F1-score | | |
|---|---|---|---|---|---|---|---|---|---|---|---|---|---|
| | | M | LQ | HQ | M | LQ | HQ | M | LQ | HQ | M | LQ | HQ |
| 0.1 | 0.3714 | 0.9047 | 0.2193 | 0 | 1 | 0.1 | 0.3714 | 0.9047 | 0.2193 | 0 | 1 | 0.1 | 0.3714 |
| 0.2 | 0.3571 | 0.8095 | 0.1938 | 0.1428 | 1 | 0.2 | 0.3571 | 0.8095 | 0.1938 | 0.1428 | 1 | 0.2 | 0.3571 |
| 0.3 | 0.3571 | 0.7857 | 0.1765 | 0.1428 | 1 | 0.3 | 0.3571 | 0.7857 | 0.1765 | 0.1428 | 1 | 0.3 | 0.3571 |
| 0.4 | 0.4142 | 0.8571 | 0.2207 | 0.4285 | 1 | 0.4 | 0.4142 | 0.8571 | 0.2207 | 0.4285 | 1 | 0.4 | 0.4142 |
| 0.5 | 0.4428 | 0.8571 | 0.2326 | 0.5714 | 1 | 0.5 | 0.4428 | 0.8571 | 0.2326 | 0.5714 | 1 | 0.5 | 0.4428 |
| 0.6 | 0.3999 | 0.7857 | 0.1802 | 0.5714 | 1 | 0.6 | 0.3999 | 0.7857 | 0.1802 | 0.5714 | 1 | 0.6 | 0.3999 |
| 0.7 | 0.4571 | 0.8095 | 0.2414 | 0.7142 | 1 | 0.7 | 0.4571 | 0.8095 | 0.2414 | 0.7142 | 1 | 0.7 | 0.4571 |
| 0.8 | 0.4285 | 0.6285 | 0.1897 | 1 | 0.8571 | 0.8 | 0.4285 | 0.6285 | 0.1897 | 1 | 0.8571 | 0.8 | 0.4285 |
| 0.9 | 0.3714 | 0.6142 | 0.1054 | 0.8571 | 0.9285 | 0.9 | 0.3714 | 0.6142 | 0.1054 | 0.8571 | 0.9285 | 0.9 | 0.3714 |

Based on the results, Range 1 outperforms the other two ranges overall. Specifically, Figure 7 shows that the specificity and recall values for Range 1 at a weight of 0.1 exceed 85%, indicating that more than 85% of ESs were accurately classified across the three statuses. Consequently, the value of α in (15) is set to 0.1. Furthermore, both the precision and accuracy metrics achieve their highest values at this weight, underscoring the model's effectiveness at this setting.

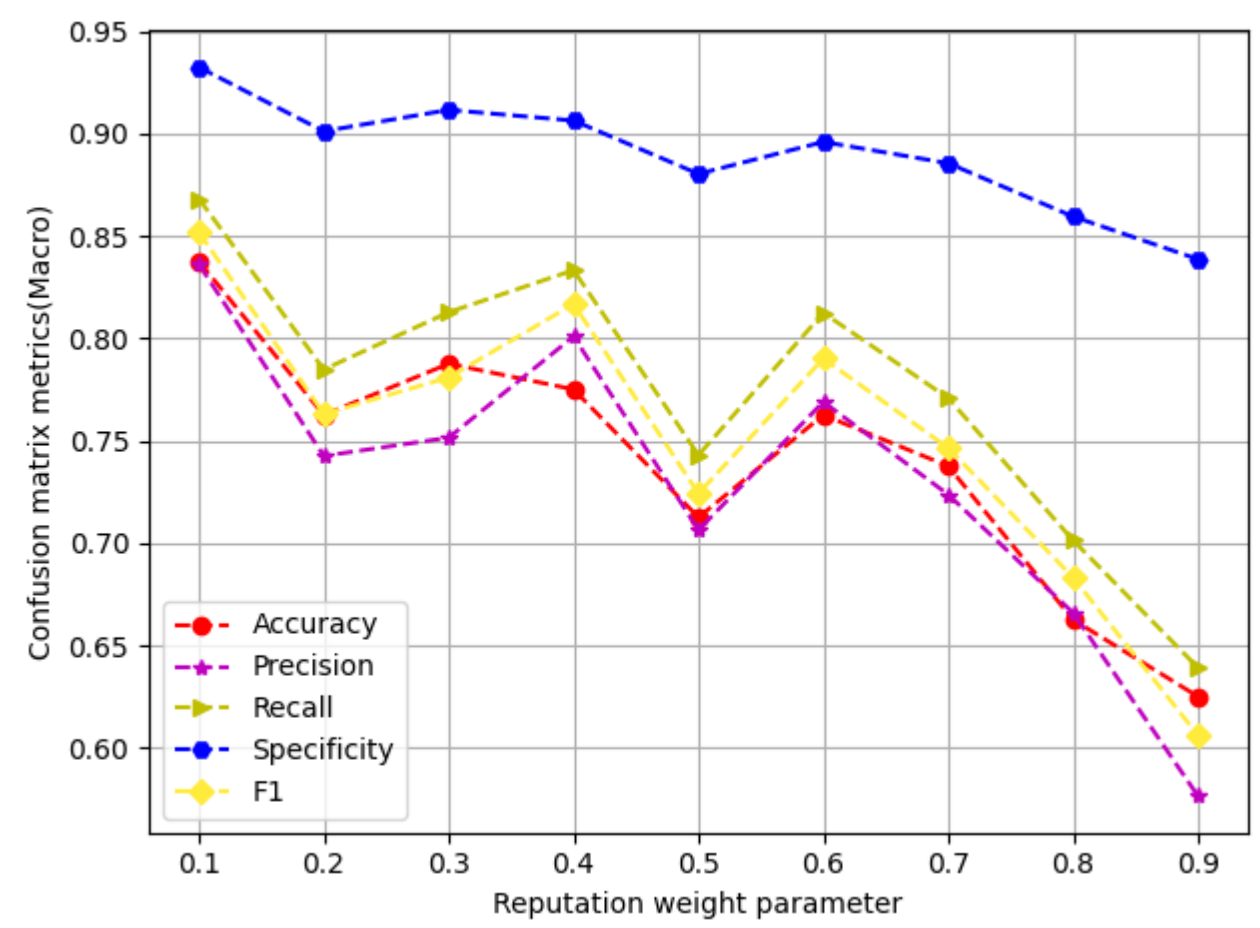

**FIGURE 7. Macro Average for Confusion Matrix Metrics in Range 1**

The trust threshold for assessing the trustworthiness of an edge server (ES) is determined using (16). Table 8 showcases the evaluation results of the proposed method, utilizing confusion matrix metrics to identify the optimal weight parameter $\beta$ in (16). Figure 10 visually represents the macro-averaged precision, recall, specificity, and F1-score derived from Table 8.

In Figure 10, the specificity and recall values exceed 0.8 at a weight of 0.8, signifying that more than 80% of ESs were accurately classified across the malicious, low-quality, and high-quality categories. This led to the selection of 0.8 as the value for $\beta$ in (16). Additionally, the precision and accuracy metrics also reach their peak at this weight, highlighting the model's enhanced performance when $\beta$ is set to 0.8.

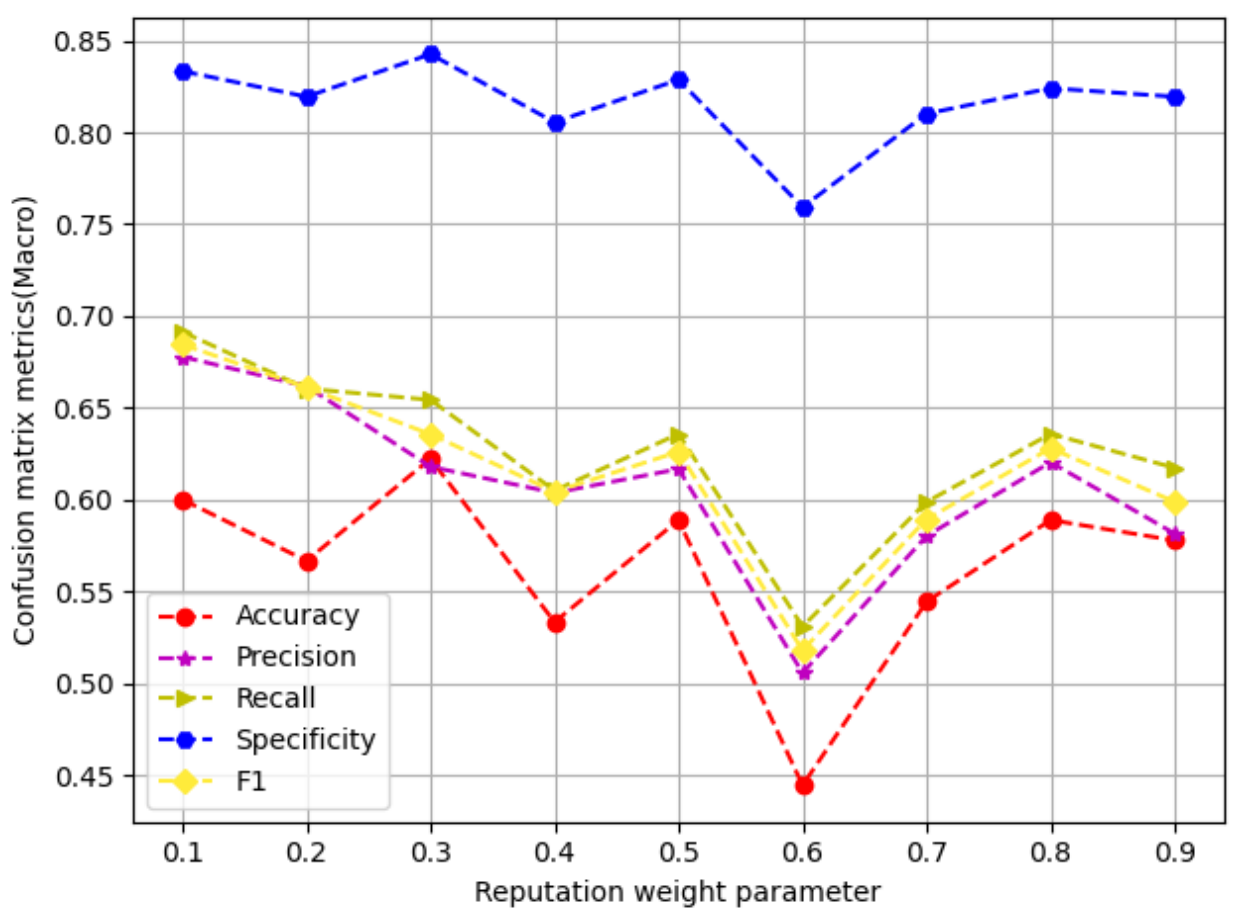

**FIGURE 8.** **Macro Average for Confusion Matrix Metrics in Range 2**

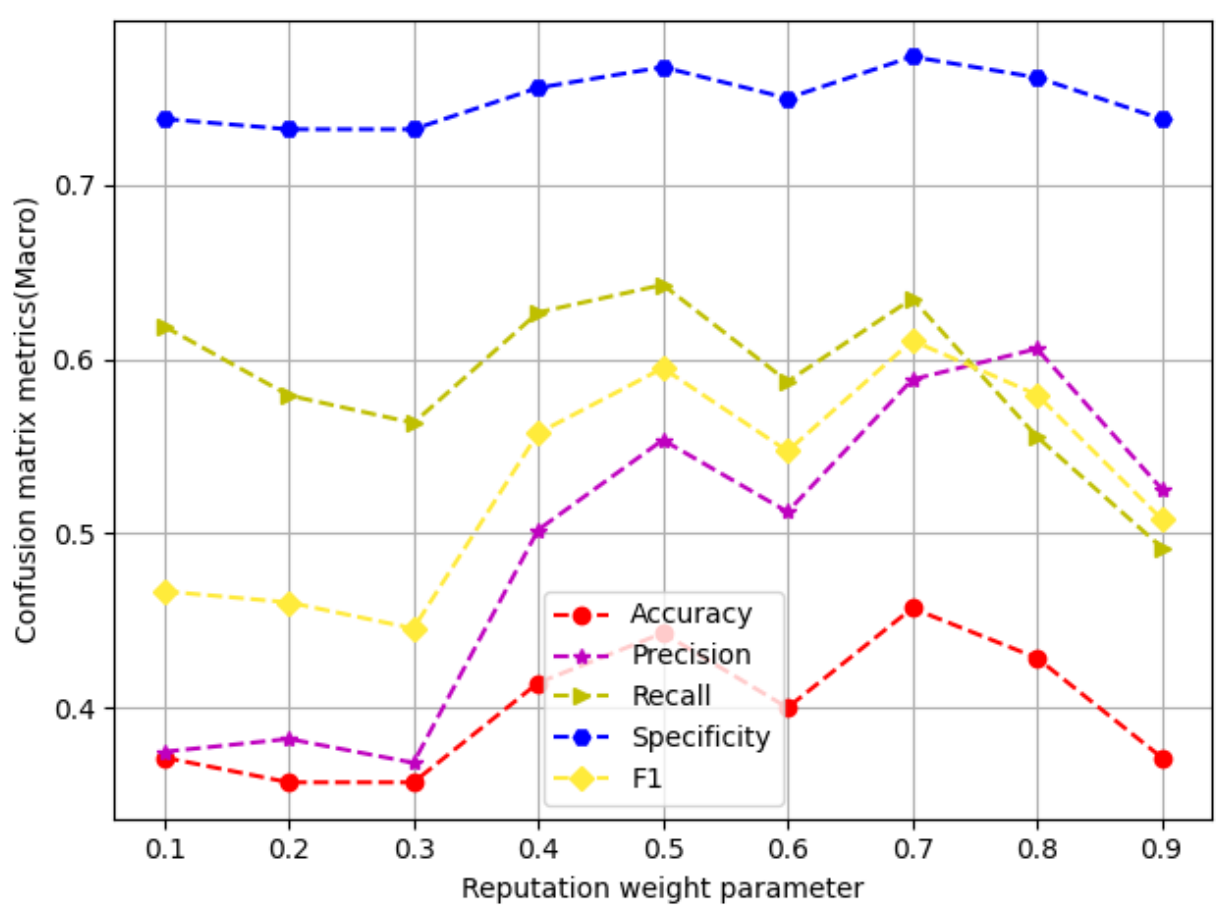

**FIGURE 9.** **Macro Average for Confusion Matrix Metrics in Range 3**

TABLE 8
TRUST THRESHOLD CONFUSION MATRIX METRICS VALUES

| $\beta$ | Accuracy | Precision | | | Recall | | | Specificity | | | F1-score | | |
|---|---|---|---|---|---|---|---|---|---|---|---|---|---|
| | | M | LQ | HQ | M | LQ | HQ | M | LQ | HQ | M | LQ | HQ |
| 0.1 | 0.7 | 0.6296 | 0.2814 | 1 | 1 | 0.1 | 0.7 | 0.6296 | 0.2814 | 1 | 1 | 0.1 | 0.7 |
| 0.2 | 0.7888 | 0.6296 | 0.5 | 1 | 1 | 0.2 | 0.7888 | 0.6296 | 0.5 | 1 | 1 | 0.2 | 0.7888 |
| 0.3 | 0.8222 | 0.7407 | 0.4444 | 1 | 1 | 0.3 | 0.8222 | 0.7407 | 0.4444 | 1 | 1 | 0.3 | 0.8222 |
| 0.4 | 0.8 | 0.6851 | 0.3888 | 1 | 1 | 0.4 | 0.8 | 0.6851 | 0.3888 | 1 | 1 | 0.4 | 0.8 |
| 0.5 | 0.7222 | 0.5925 | 0.3888 | 1 | 1 | 0.5 | 0.7222 | 0.5925 | 0.3888 | 1 | 1 | 0.5 | 0.7222 |
| 0.6 | 0.7777 | 0.6481 | 0.4444 | 1 | 1 | 0.6 | 0.7777 | 0.6481 | 0.4444 | 1 | 1 | 0.6 | 0.7777 |
| 0.7 | 0.8 | 0.6666 | 0.5092 | 1 | 1 | 0.7 | 0.8 | 0.6666 | 0.5092 | 1 | 1 | 0.7 | 0.8 |
| 0.8 | 0.8111 | 0.7777 | 0.5259 | 1 | 1 | 0.8 | 0.8111 | 0.7777 | 0.5259 | 1 | 1 | 0.8 | 0.8111 |
| 0.9 | 0.7333 | 0.6111 | 0.2685 | 1 | 1 | 0.9 | 0.7333 | 0.6111 | 0.2685 | 1 | 1 | 0.9 | 0.7333 |

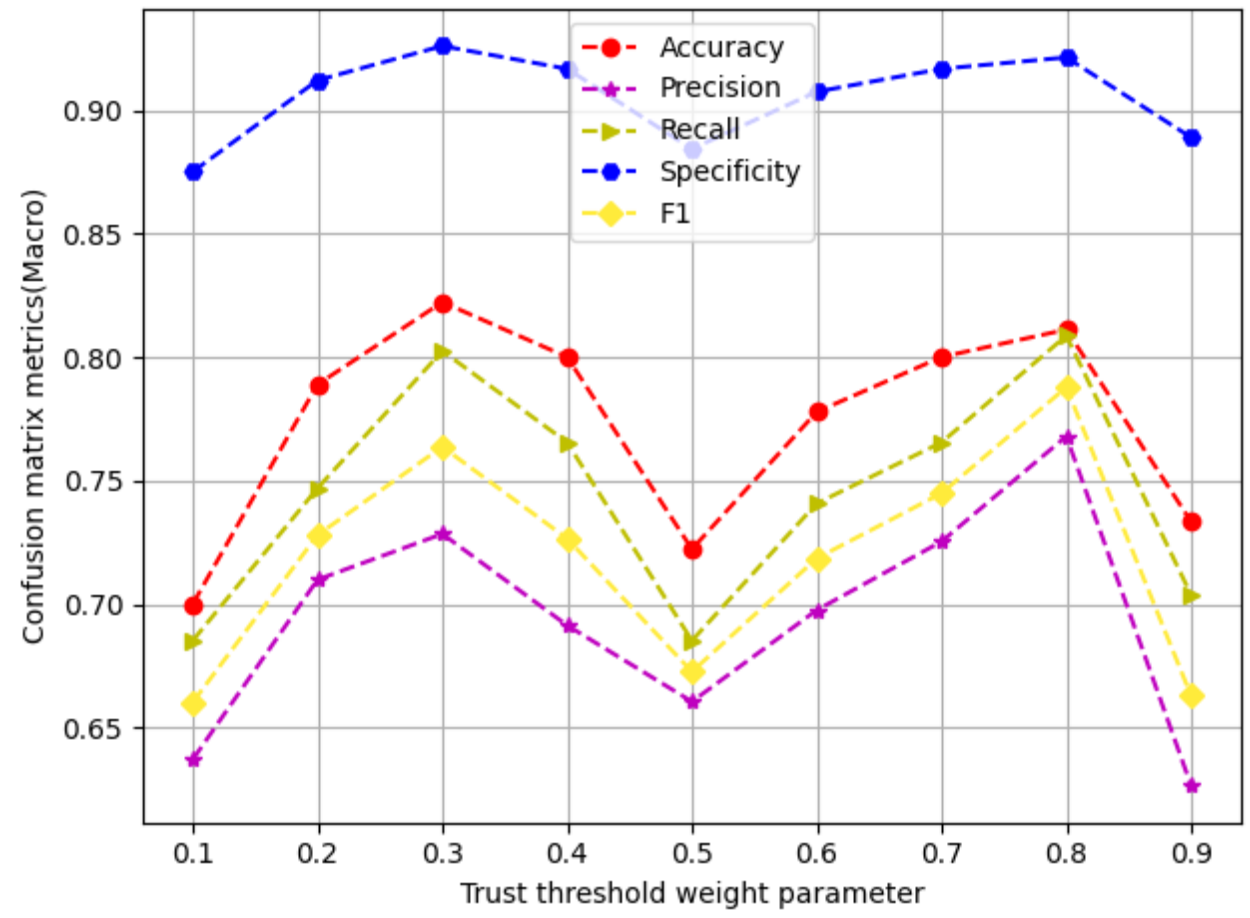

**FIGURE 10.** **Macro Average for Trust Threshold Confusion Matrix Metrics**

The weight parameter ω for determining the cache status of an edge server (ES) is calculated using (18). Table 9 presents the evaluation results based on confusion matrix metrics to identify the optimal value for ω. Figure 11 illustrates the macro-averaged metrics of precision, recall, specificity, and F1-score derived from Table 9.

TABLE 9
EDGE SERVER CACHE STATUS CONFUSION MATRIX METRICS VALUES

| $\omega$ | Accuracy | Precision | | | Recall | | | Specificity | | | F1-score | | |
|---|---|---|---|---|---|---|---|---|---|---|---|---|---|
| | | M | LQ | HQ | M | LQ | HQ | M | LQ | HQ | M | LQ | HQ |
| 0.1 | 0.7125 | 0.6666 | 0.2916 | 0.9375 | 1 | 0.1 | 0.7125 | 0.6666 | 0.2916 | 0.9375 | 1 | 0.1 | 0.7125 |
| 0.2 | 0.7374 | 0.6875 | 0.3645 | 1 | 1 | 0.2 | 0.7374 | 0.6875 | 0.3645 | 1 | 1 | 0.2 | 0.7374 |
| 0.3 | 0.7374 | 0.7708 | 0.3729 | 0.975 | 1 | 0.3 | 0.7374 | 0.7708 | 0.3729 | 0.975 | 1 | 0.3 | 0.7374 |
| 0.4 | 0.8 | 0.7708 | 0.5 | 1 | 1 | 0.4 | 0.8 | 0.7708 | 0.5 | 1 | 1 | 0.4 | 0.8 |
| 0.5 | 0.6875 | 0.75 | 0.302 | 0.9187 | 1 | 0.5 | 0.6875 | 0.75 | 0.302 | 0.9187 | 1 | 0.5 | 0.6875 |
| 0.6 | 0.6625 | 0.6875 | 0.3125 | 0.9583 | 0.9375 | 0.6 | 0.6625 | 0.6875 | 0.3125 | 0.9583 | 0.9375 | 0.6 | 0.6625 |
| 0.7 | 0.7125 | 0.8125 | 0.4562 | 0.9437 | 0.9375 | 0.7 | 0.7125 | 0.8125 | 0.4562 | 0.9437 | 0.9375 | 0.7 | 0.7125 |
| 0.8 | 0.6625 | 0.7291 | 0.2354 | 0.975 | 0.9375 | 0.8 | 0.6625 | 0.7291 | 0.2354 | 0.975 | 0.9375 | 0.8 | 0.6625 |
| 0.9 | 0.6249 | 0.7708 | 0.2812 | 0.8437 | 0.875 | 0.9 | 0.6249 | 0.7708 | 0.2812 | 0.8437 | 0.875 | 0.9 | 0.6249 |

As depicted in Figure 11, the specificity and recall values at a weight of 0.4 exceed 0.8, indicating that more than 80% of ESs were correctly classified into the three categories: malicious, low-quality, and high-quality. Consequently, the weight parameter $\omega$ in (18) is set to 0.4. Moreover, precision and accuracy metrics also reach their highest values at this weight, further emphasizing the model's optimal performance when $\omega$ is 0.4.

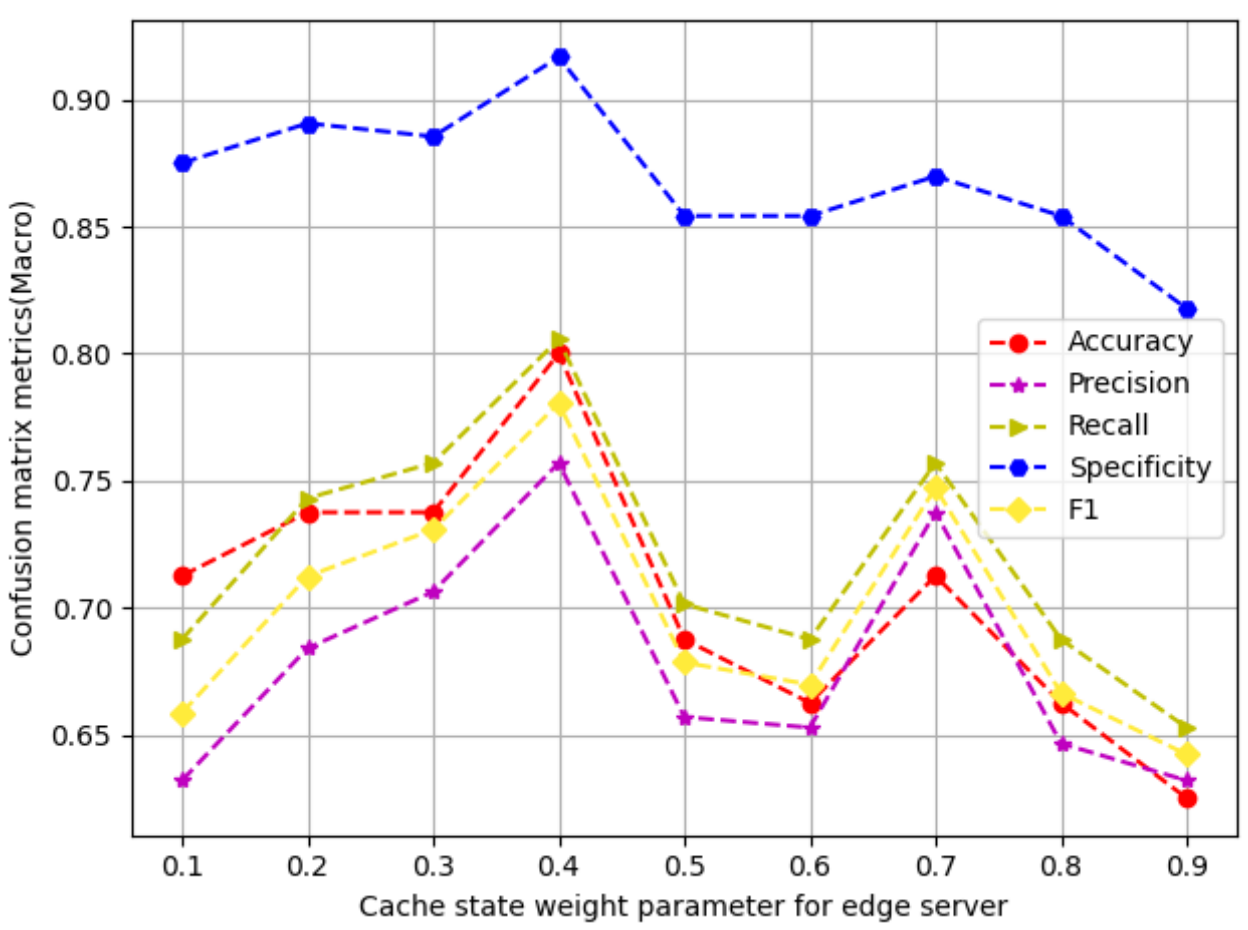

**FIGURE 11. Macro Average for Edge Server Cache Status Confusion Matrix Metrics**

The weight parameter $\tilde{\alpha}$ for determining the cache status of a cooperative mobile node is calculated using (20). Table 10 presents the evaluation results based on confusion matrix metrics to identify the optimal value for $\tilde{\alpha}$. The macro-averaged precision, recall, specificity, and F1-scores derived from Table 10 are illustrated in Figure 12.

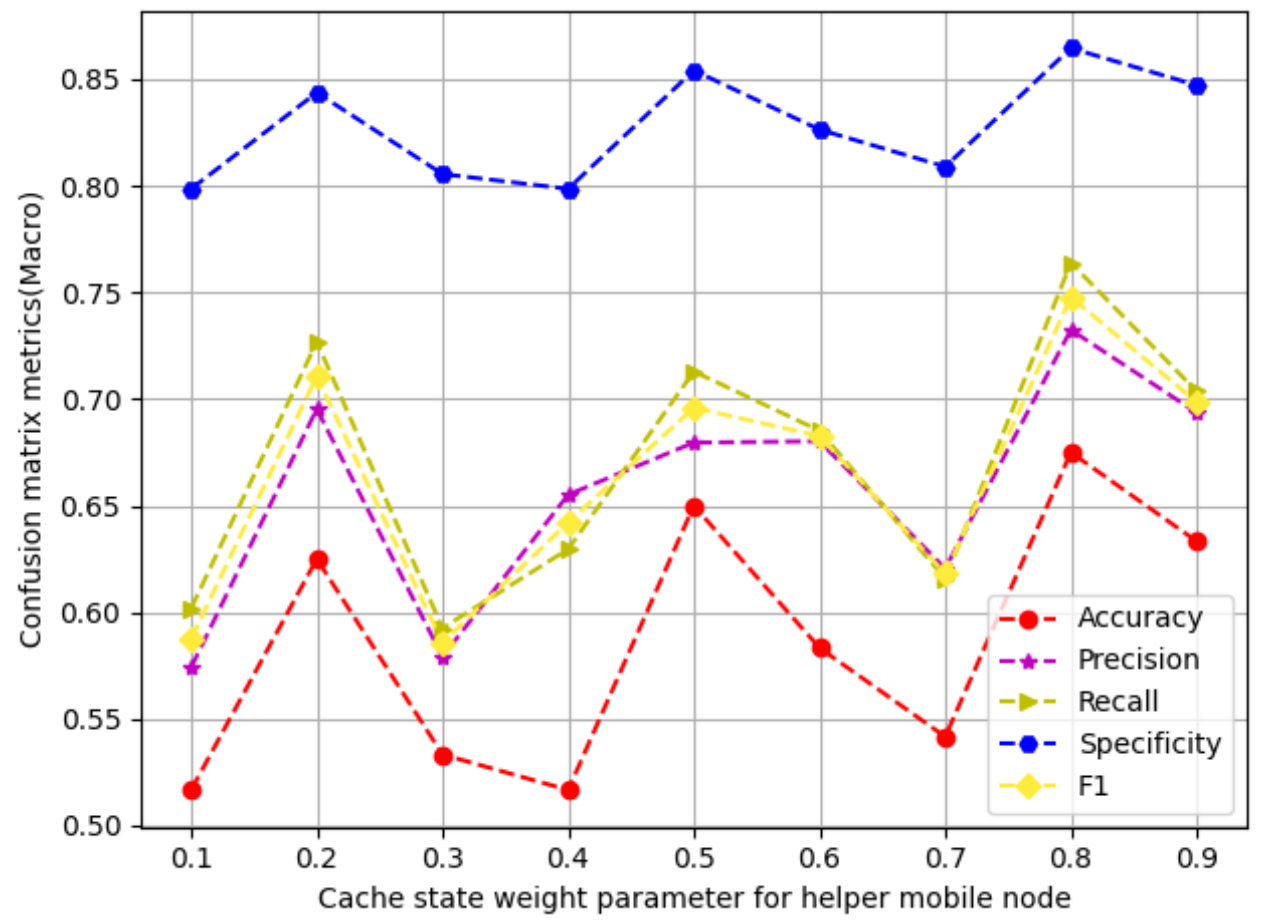

**FIGURE 12. Macro Average for Cooperative Mobile Node Cache Status Confusion Matrix Metrics**

As shown in Figure 12, the specificity and recall values at a weight of 0.8 exceed 0.75, indicating that over 75% of cooperative mobile nodes were correctly classified into the categories of malicious (M), low-quality (LQ), and high-quality (HQ). Consequently, the weight parameter $\tilde{\alpha}$ in (20) is determined to be 0.8. Furthermore, precision and accuracy metrics also reach their peak at this weight, highlighting the model's optimal performance when $\tilde{\alpha}$ is set to 0.8.

TABLE 10
COOPERATIVE MOBILE NODE CACHE STATUS CONFUSION MATRIX METRICS VALUES

| $\tilde{\alpha}$ | Accuracy | Precision | | | Recall | | | Specificity | | | F1-score | | |
|---|---|---|---|---|---|---|---|---|---|---|---|---|---|
| | | M | LQ | HQ | M | LQ | HQ | M | LQ | HQ | M | LQ | HQ |
| 0.1 | 0.5166 | 0.6083 | 0.1974 | 0.9166 | 0.9583 | 0.1 | 0.5166 | 0.6083 | 0.1974 | 0.9166 | 0.9583 | 0.1 | 0.5166 |
| 0.2 | 0.625 | 0.7361 | 0.3499 | 1 | 1 | 0.2 | 0.625 | 0.7361 | 0.3499 | 1 | 1 | 0.2 | 0.625 |
| 0.3 | 0.5333 | 0.5861 | 0.1513 | 1 | 1 | 0.3 | 0.5333 | 0.5861 | 0.1513 | 1 | 1 | 0.3 | 0.5333 |
| 0.4 | 0.5166 | 0.7083 | 0.2571 | 1 | 0.9166 | 0.4 | 0.5166 | 0.7083 | 0.2571 | 1 | 0.9166 | 0.4 | 0.5166 |
| 0.5 | 0.65 | 0.7361 | 0.3027 | 1 | 1 | 0.5 | 0.65 | 0.7361 | 0.3027 | 1 | 1 | 0.5 | 0.65 |
| 0.6 | 0.5833 | 0.7222 | 0.3184 | 1 | 1 | 0.6 | 0.5833 | 0.7222 | 0.3184 | 1 | 1 | 0.6 | 0.5833 |
| 0.7 | 0.5416 | 0.5888 | 0.2726 | 1 | 0.7916 | 0.7 | 0.5416 | 0.5888 | 0.2726 | 1 | 0.7916 | 0.7 | 0.5416 |
| 0.8 | 0.6750 | 0.7638 | 0.4335 | 1 | 1 | 0.8 | 0.6750 | 0.7638 | 0.4335 | 1 | 1 | 0.8 | 0.6750 |
| 0.9 | 0.6333 | 0.7361 | 0.3448 | 1 | 0.9166 | 0.9 | 0.6333 | 0.7361 | 0.3448 | 1 | 0.9166 | 0.9 | 0.6333 |

### *B. Evaluation of the Proposed Method*

In this section, we present a comprehensive analysis of the proposed method's performance across various edge server scenarios. The evaluation focuses on critical aspects such as pricing dynamics, trust stability, scalability, and resilience to noise, all of which are essential for effective trust management in edge computing environments. Through a series of simulations, we examine how the model adapts to different proportions of low-quality servers, increasing user demand, and potential data inconsistencies. The results demonstrate that the proposed approach maintains stable trust levels and accurate classification of edge servers, even as network conditions fluctuate. This analysis provides insights into the model's capacity to enhance security, optimize resource allocation, and ensure reliable service delivery in dynamic edge networks.

#### 1) EFFECTIVENESS OF THE INCENTIVE MECHANISM:

To evaluate the effect of the incentive mechanism on cache sharing, we simulated a scenario in which edge servers experienced cache shortages and relied on cooperative mobile nodes to compensate. The results showed that increasing the trust-weighted reward (Equation 22) led to a significant increase in the number of participating caching mobile users (CMUs). Specifically, nodes with higher trust values were more likely to contribute cache resources, as the reward mechanism provided proportional token-based compensation. This incentivized behavior improved overall content availability and reduced delivery latency. These results confirm that the incentive design effectively encourages cooperation in both collaborative and partially competitive environments.

It is worth noting that while the incentive mechanism is designed to effectively encourage cooperation, its actual impact on network performance may vary depending on specific network conditions and user behaviors. Therefore, further empirical evaluation under diverse scenarios is necessary to fully understand and optimize its effectiveness.

Unlike conventional content-caching approaches that lack trust-driven incentive mechanisms, our proposed system employs dynamic, trust-based reward adjustments to align user behavior with network objectives. This innovation fosters increased cooperation and more efficient resource sharing, thereby enhancing overall network performance.
The incentive function and its integration with the trust model are original contributions of this work, designed specifically to promote cooperation in edge computing content caching using blockchain-based trust computation.

### 2) SCALABILITY: PRICING BY SERVER QUALITY

Based on the obtained results and the parameter values determined, the proposed method is evaluated. Figures 13, 14, 15 and 16 illustrate the price of ESs with an increase in the percentage of low-quality ESs compared to the total number of ESs. In these figures, the proportion of low-quality edge servers is set at 20%, 40%, 60%, and 80%, respectively. An experiment was conducted with 20 ESs and one MU having a demand of 2000 across 200 interactions. The MU's demand was divided into 200 equal parts, and in each interaction, one part of the demand was sent to the optimally selected ES.

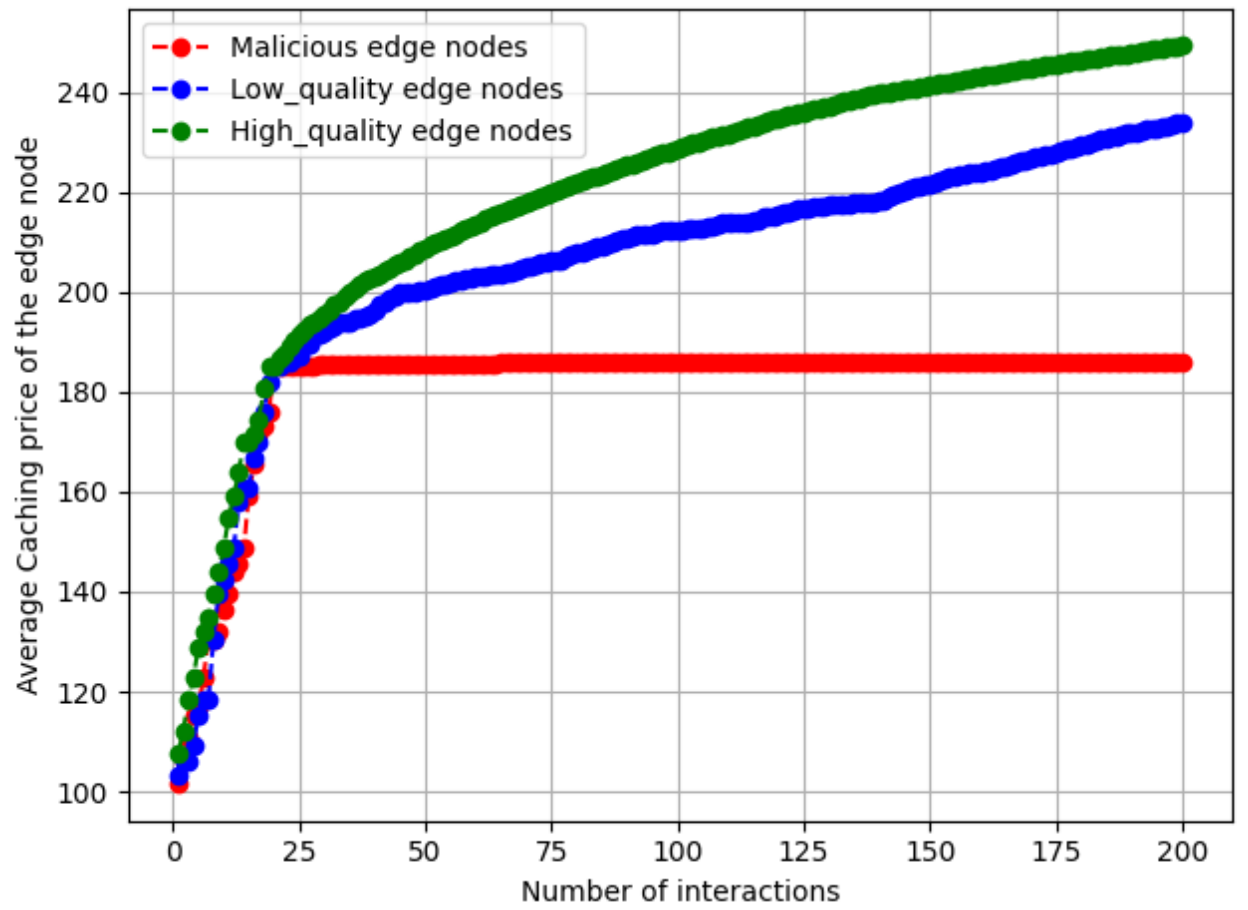

**FIGURE 13.** The caching price with 20% Low-Quality Edge Servers.

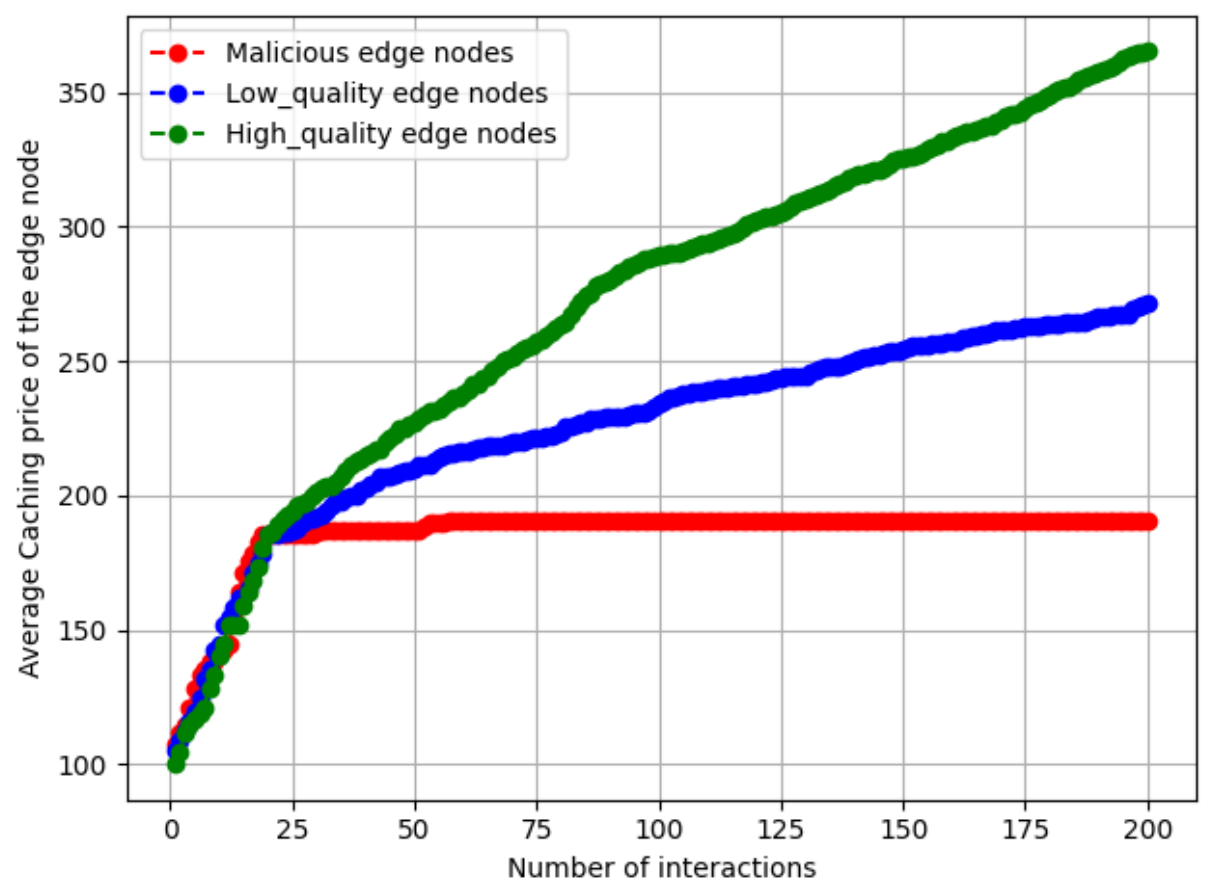

**FIGURE 14.** The caching price with 40% Low-Quality Edge Servers.

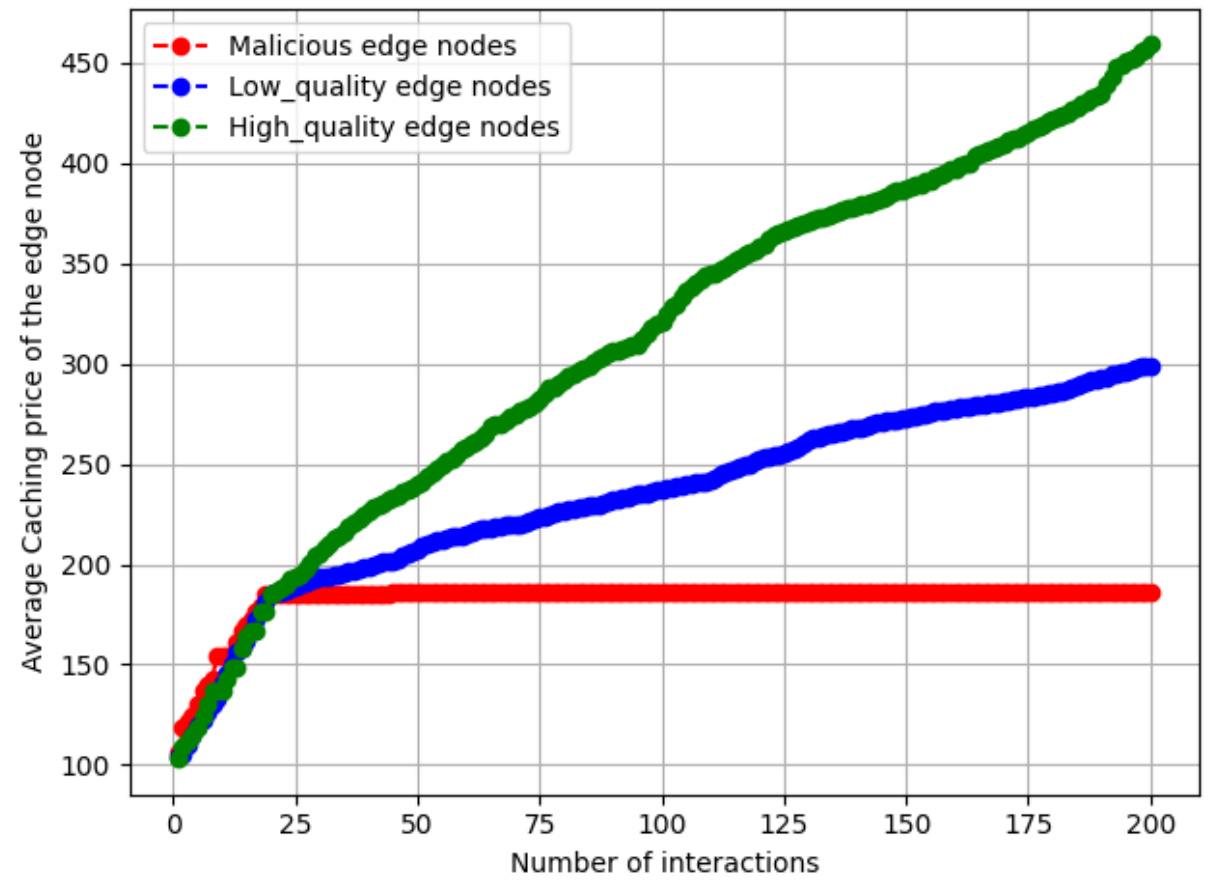

**FIGURE 15.** The caching price with 60% Low-Quality Edge Servers.

Figures 13 to 16 demonstrate that increasing the percentage of low-quality ESs leads to an increase in the price of high-quality ESs. Initially, low-quality and malicious ESs attract high demand due to their lower prices. However, as these servers provide poorer service, their demand diminishes over time, shifting towards high-quality ESs. With the increase in the percentage of low-quality servers, the number of high-quality servers decreases. As a result, when the demand for malicious and low-quality servers decline, the demand for each high-quality server increases, leading to a greater price surge for high-quality servers as the percentage of low-quality servers rises.

### 3) SCALABILITY: PRICING BY MOBILE USER COUNT

Figures 17, 18, 19 and 20 illustrate the effect of an increasing number of mobile users (MUs) on the price of edge servers (ESs). In this scenario, the experiment assumes 10 ESs, with the following distribution: 20% malicious, 60% low-quality, and 20% high-quality. Each ES has a capacity of 1000, and each MU has a demand of 50 units, which is divided into 5 equal parts. During each interaction, an MU is randomly selected to connect to the optimal ES, sequentially sending a portion of its request to that server.

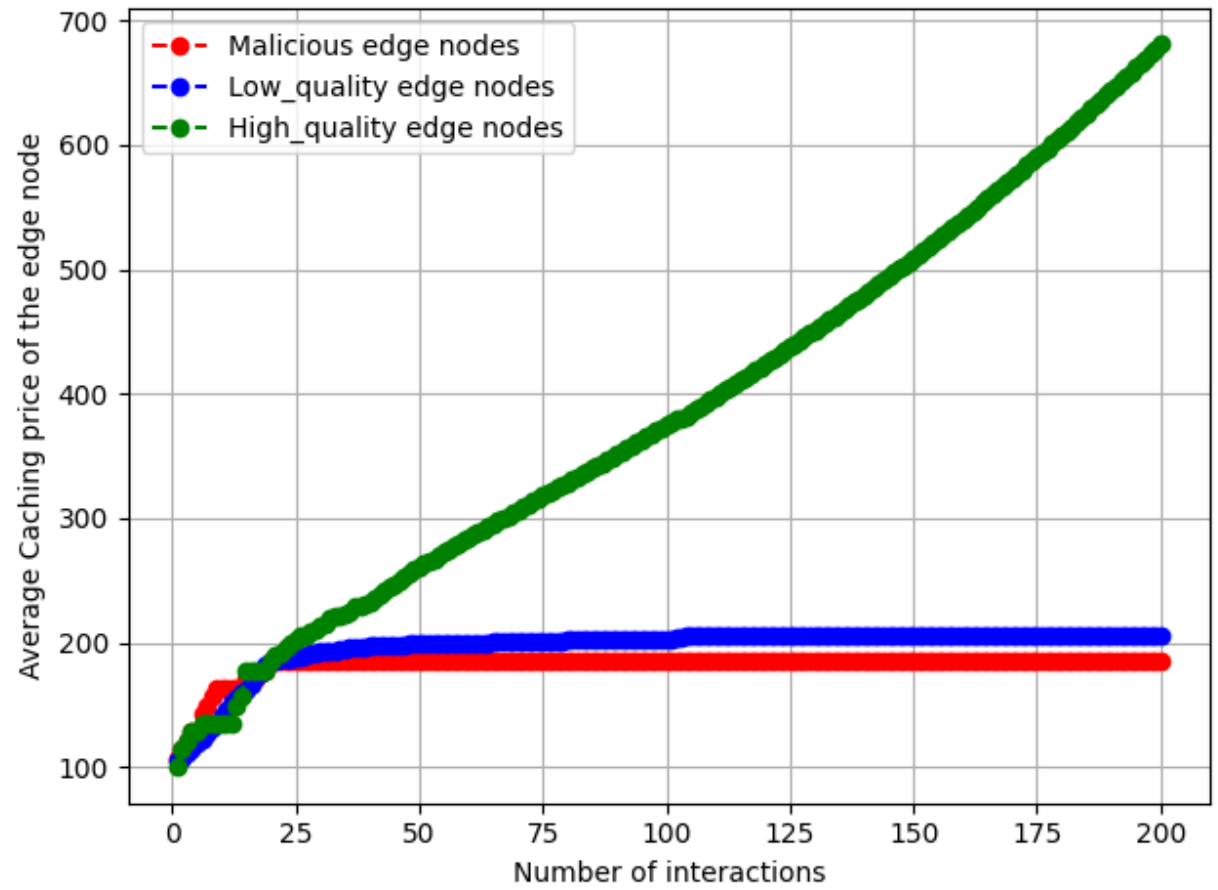

**FIGURE 16.** The caching price with 80% Low-Quality Edge Servers.

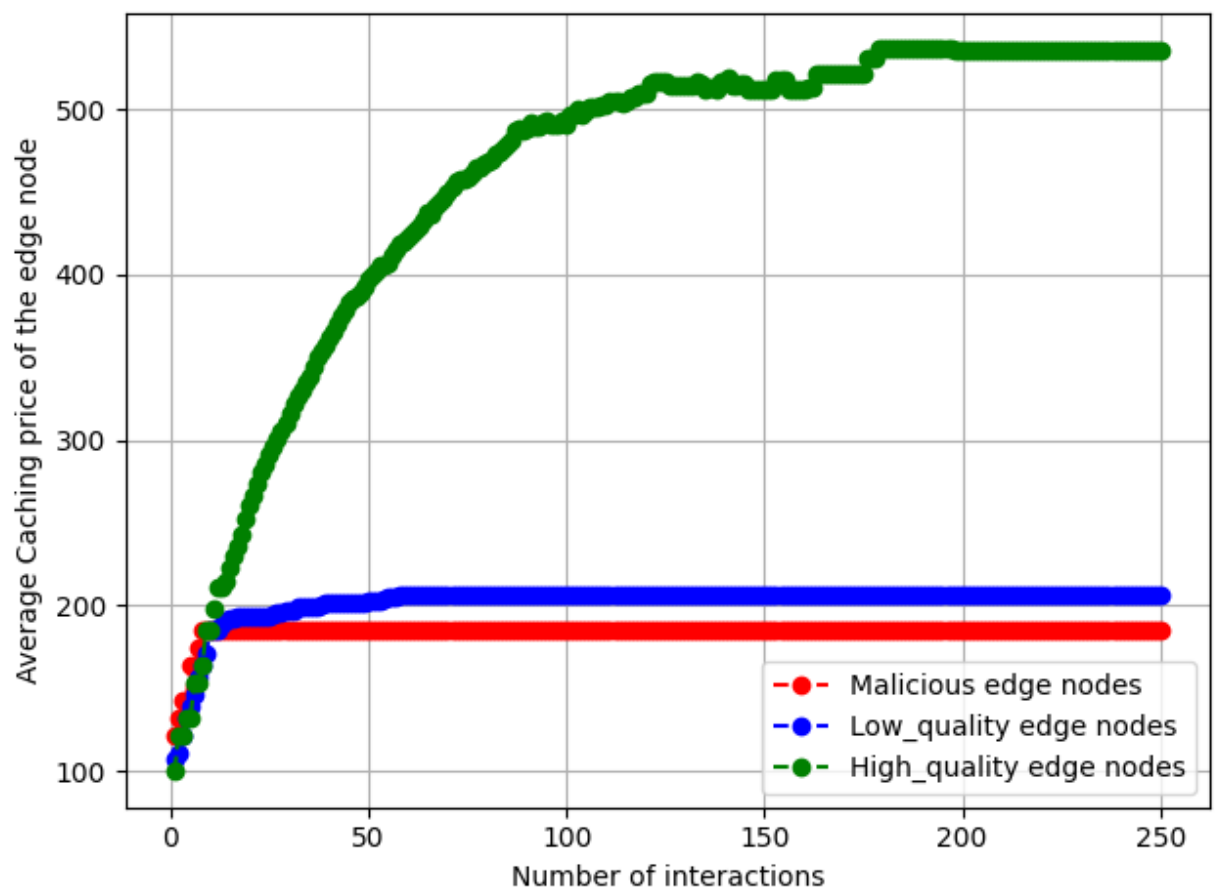

**FIGURE 17.** The Edge-Server price with 20 Mobile Users.

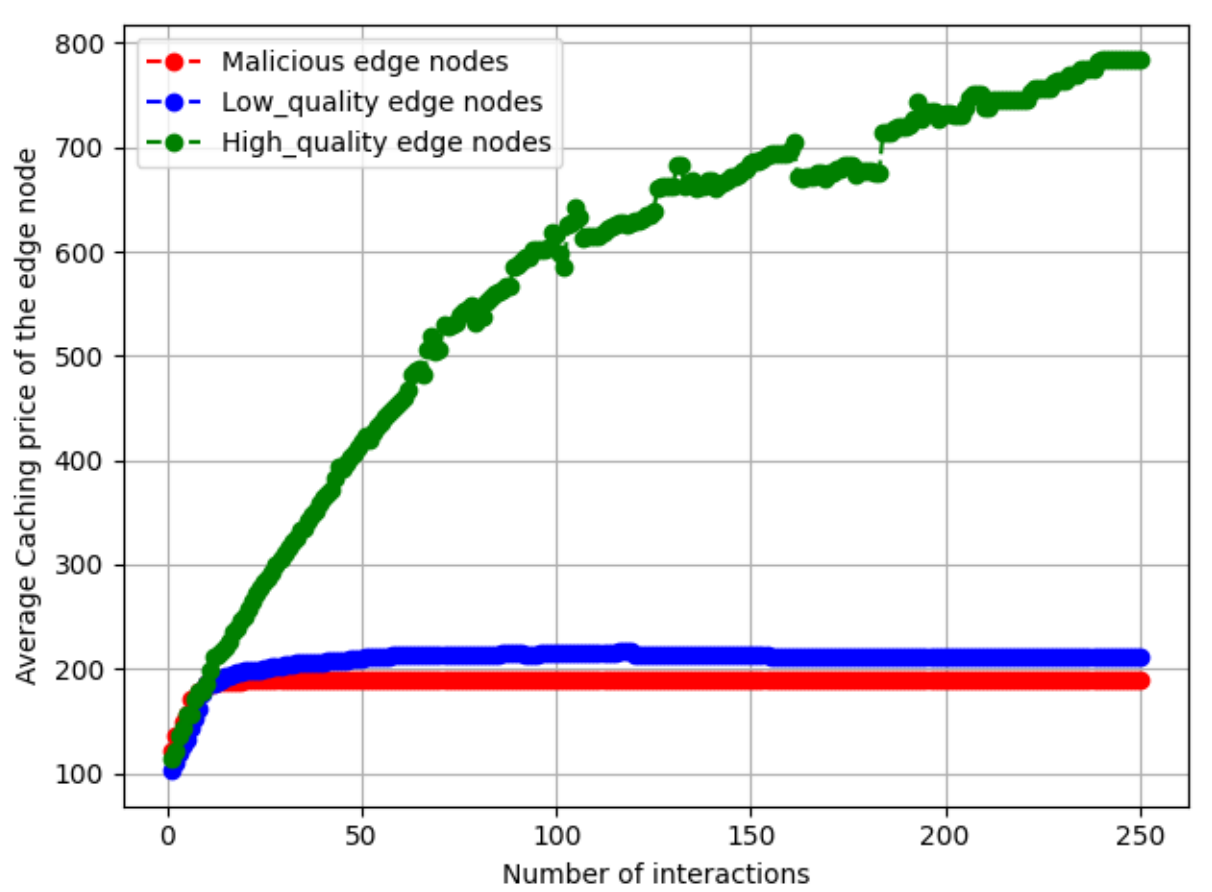

**FIGURE 18.** The Edge-Server price with 30 Mobile Users.

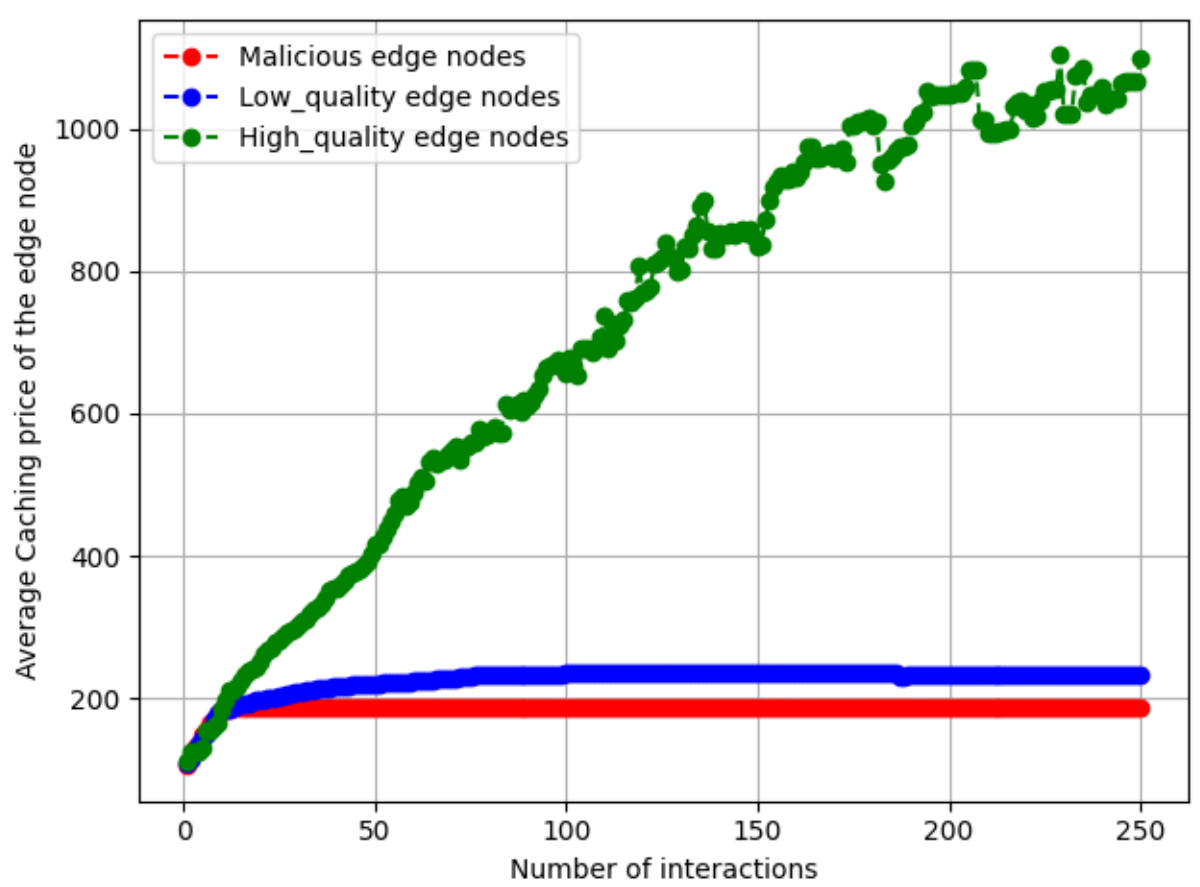

**FIGURE 19.** The Edge-Server price with 40 Mobile Users.

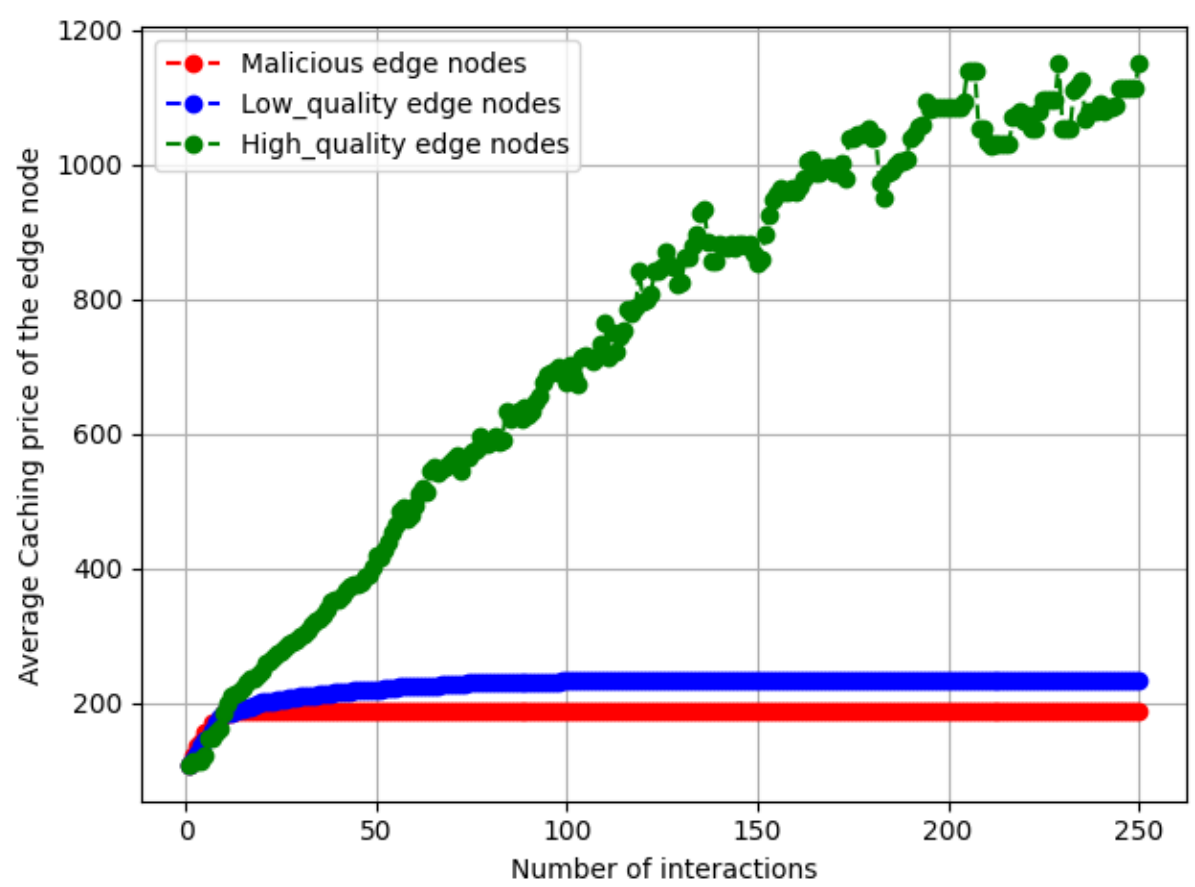

**FIGURE 20.** The Edge-Server price with 50 Mobile Users.

The simulation results show that as the number of MUs grows, the price of high-quality ESs increases. Initially, MUs tend to choose low-quality and malicious ESs because of their lower prices. However, as these servers continue to provide subpar services, their demand diminishes, leading to a shift in preference towards high-quality ESs. The increase in the number of users boosts overall demand, driving up the prices of high-quality ESs.

Unlike static pricing schemes that do not adjust based on demand or service quality, our model dynamically reflects user behavior and network conditions, resulting in more efficient resource distribution.

#### 4) TRUST ADJUSTMENTS BY SERVER QUALITY

Figures 21, 22, 23 and 24 illustrate that the trust level of high-quality ESs increases as the percentage of low-quality servers rises. This occurs because the MU interacts with fewer high-quality servers as the percentage of low-quality ESs increases. Consequently, with a reduced number of high-quality servers, the number of interactions the MU has with these servers increases. This leads to an increase in the SIC, resulting in an increase in the $f$ parameter in Equation (12) and, consequently, the trust value in Equation (15).

The model addresses collusion by relying more on indirect trust (Equation 15), which aggregates evaluations across diverse sources. Colluding nodes often generate similar biased feedback, which is flagged as low-credibility via Equation (9). Additionally, feedback that lacks corroboration from independent edge servers receives lower weight. While this method provides resilience against small-scale collusion, future work will include applying anomaly detection techniques and graph-based analysis to capture larger coordinated attacks.

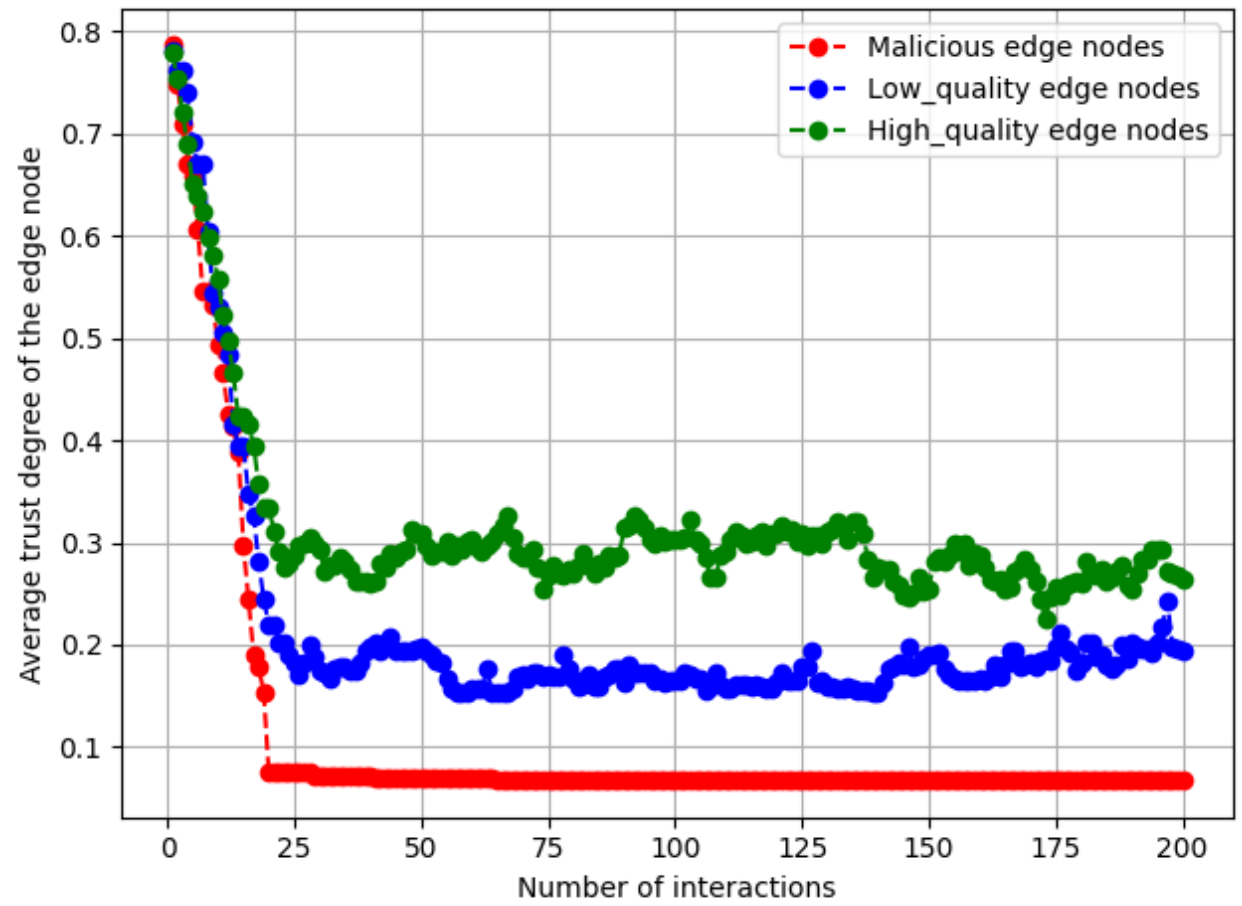

**FIGURE 21.** The edge nodes trust level with 20% Low-Quality Edge Servers

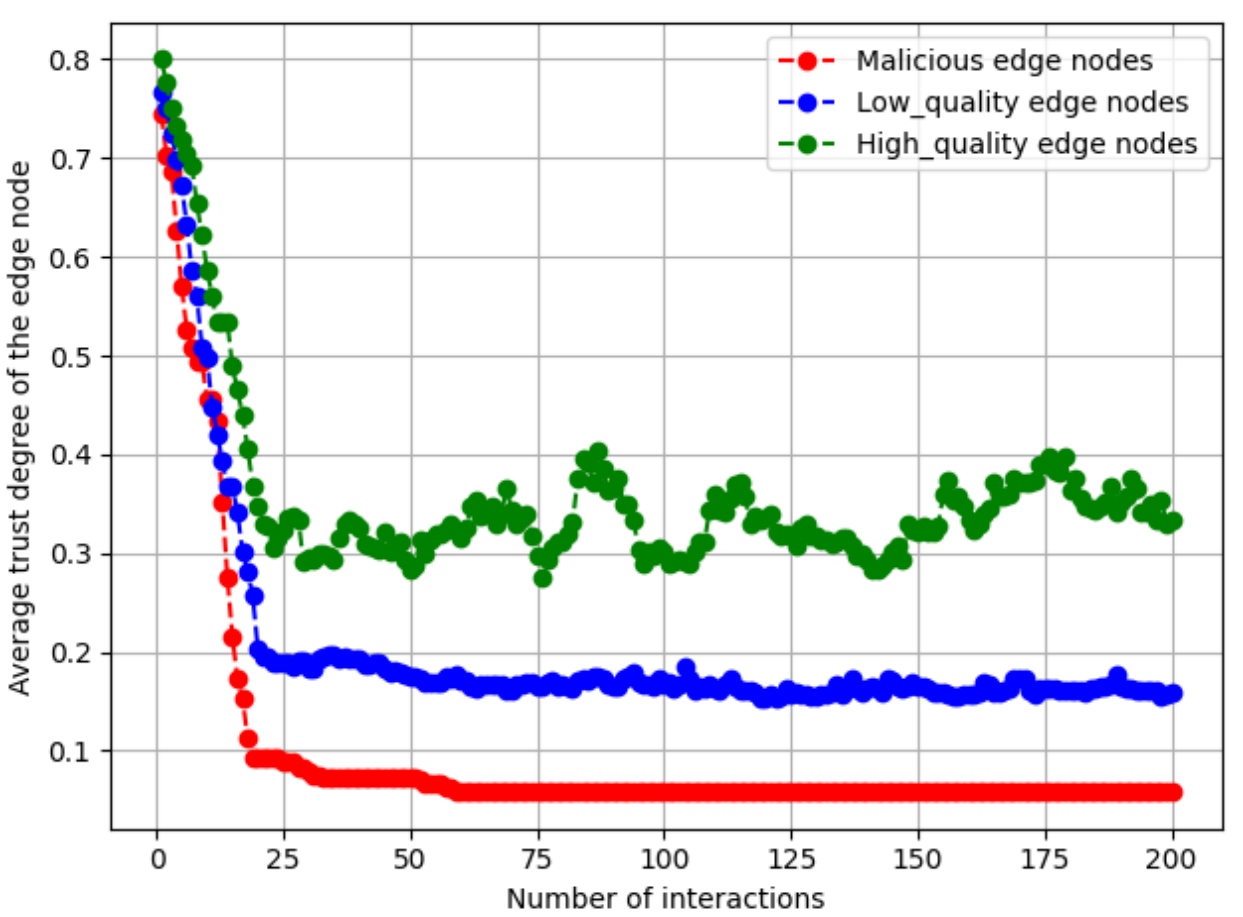

**FIGURE 22.** The edge nodes trust level with 40% Low-Quality Edge Servers

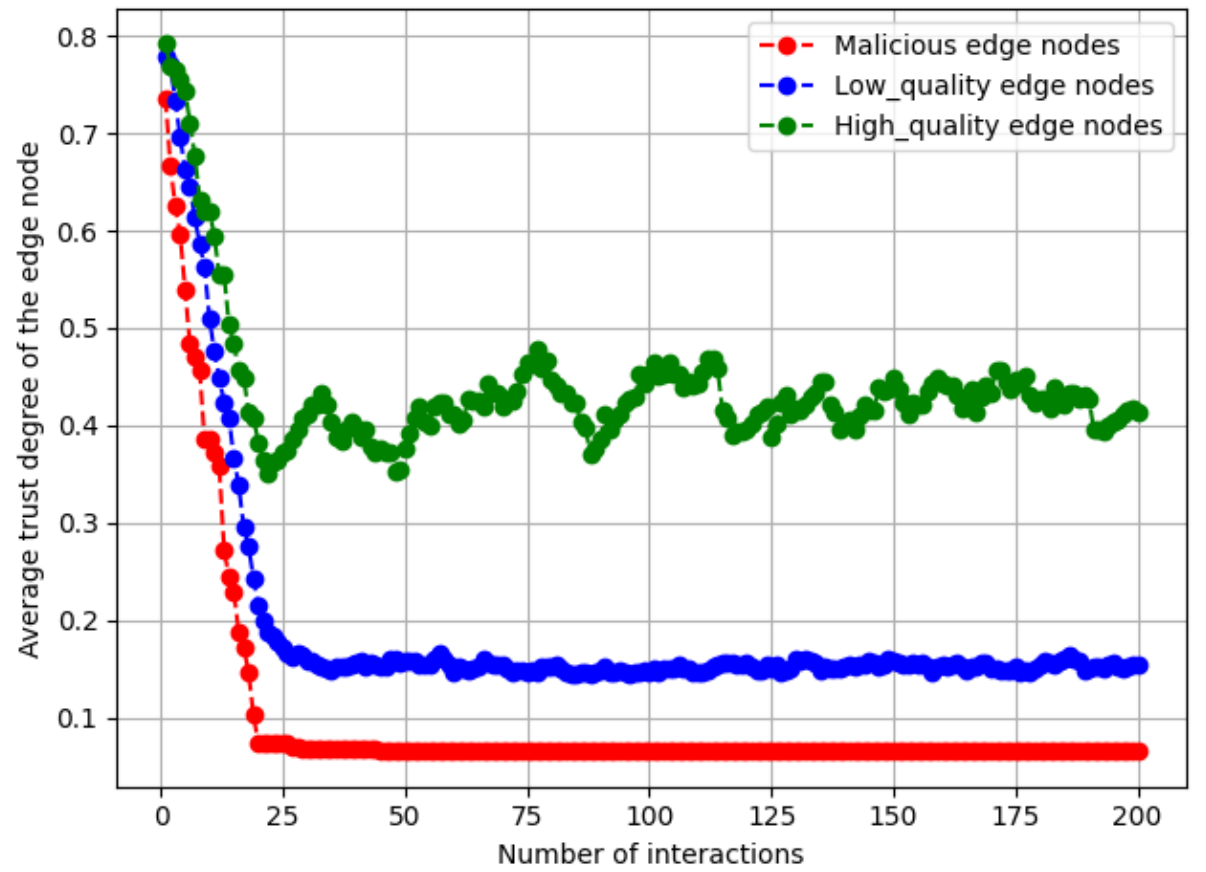

**FIGURE 23.** The edge nodes trust level with 60% Low-Quality Edge Servers

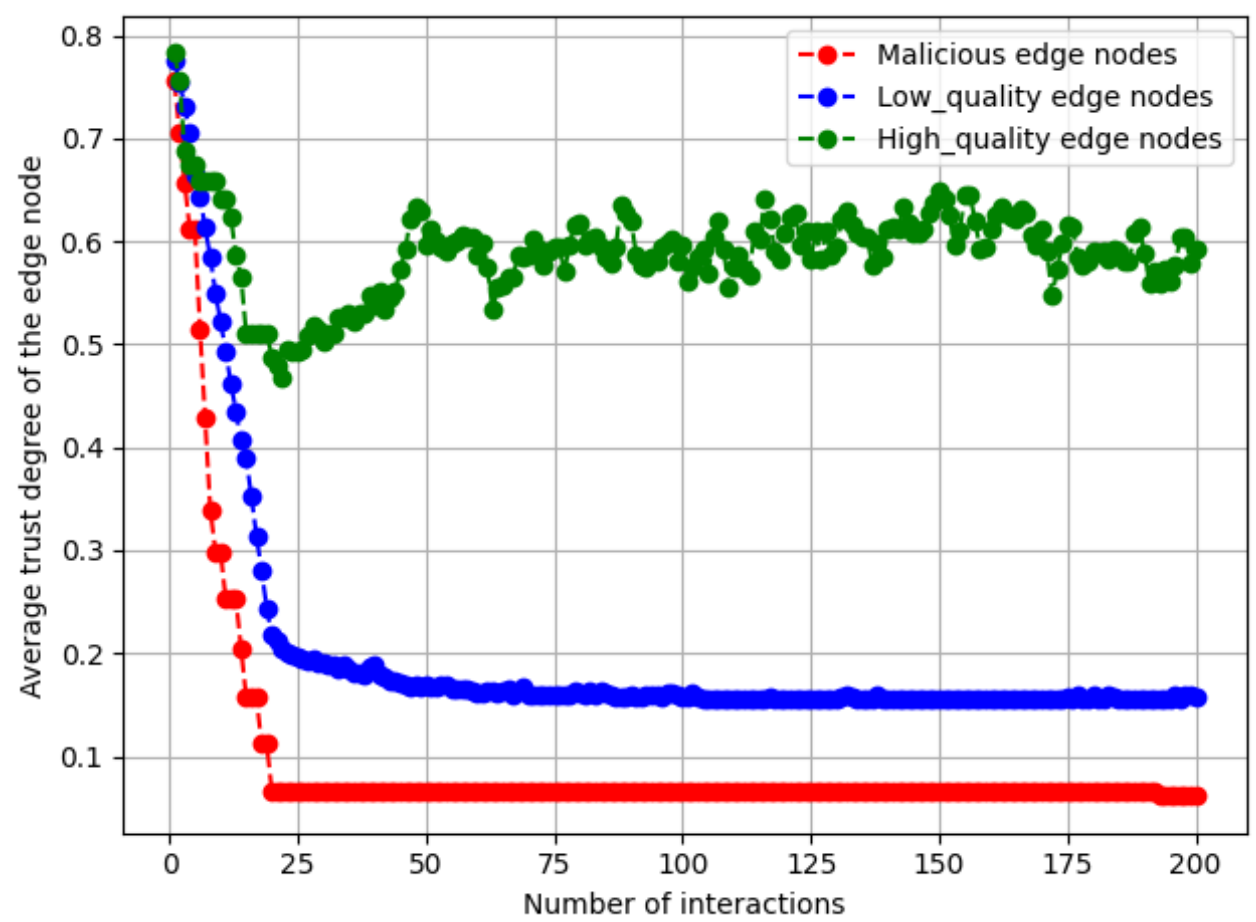

**FIGURE 24.** The edge nodes trust level with 80% Low-Quality Edge Servers.

### 5) TRUST ADJUSTMENTS BY MOBILE USER COUNT

As depicted in Figures 25, 26, 27, and 28, the trust levels of ESs remain relatively stable despite the growing number of users. This stability is due to the fact that with more users, the SIC (successful interaction count) of each MU with an ES decreases. This reduction leads to a lower value of $f$ in Equation (12), which prevents an increase in the trust values of ESs as per (15). The correction parameter $f$ plays a crucial role in ensuring that malicious ESs cannot gain high trust levels from a limited number of successful interactions, and ensures that only consistently high-quality service providers can achieve higher trust scores.

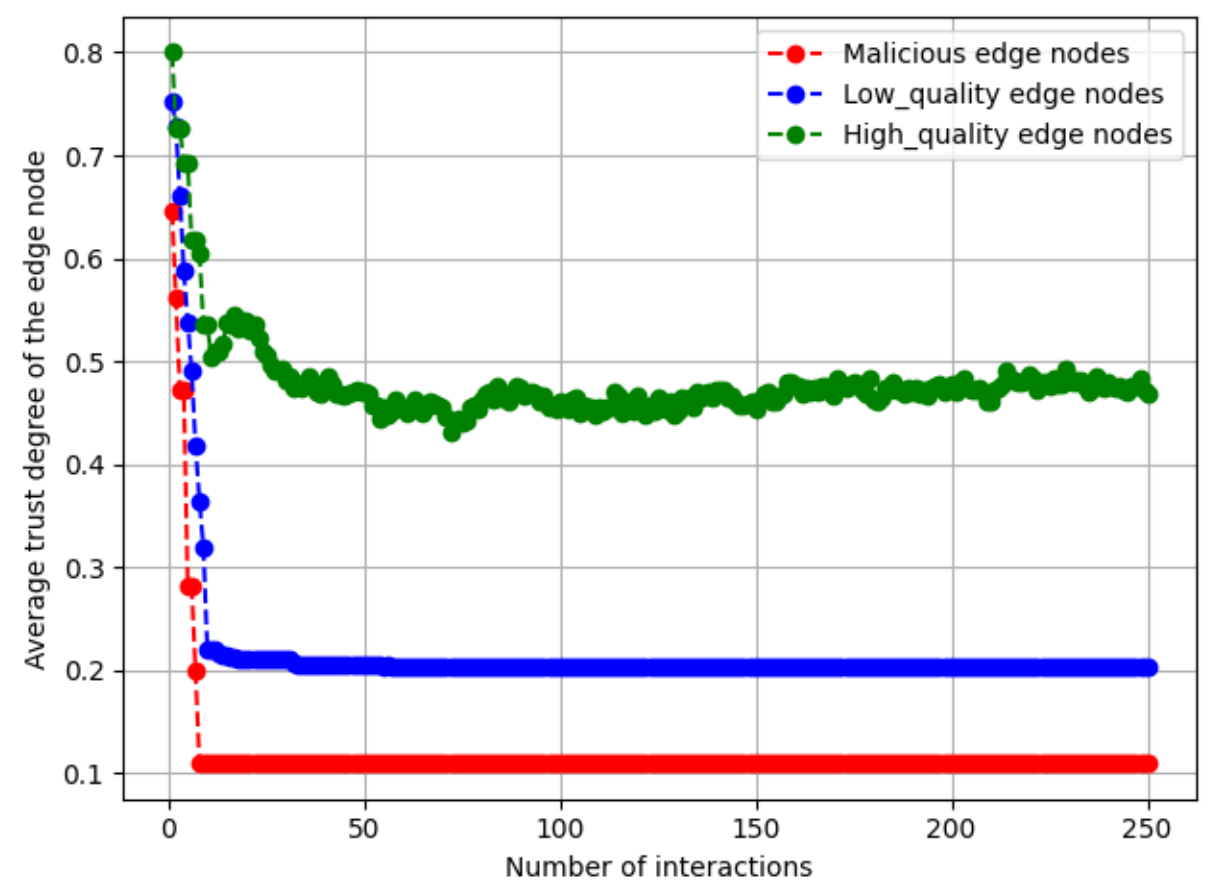

**FIGURE 25.** The edge nodes trust level with 20 Mobile Users.

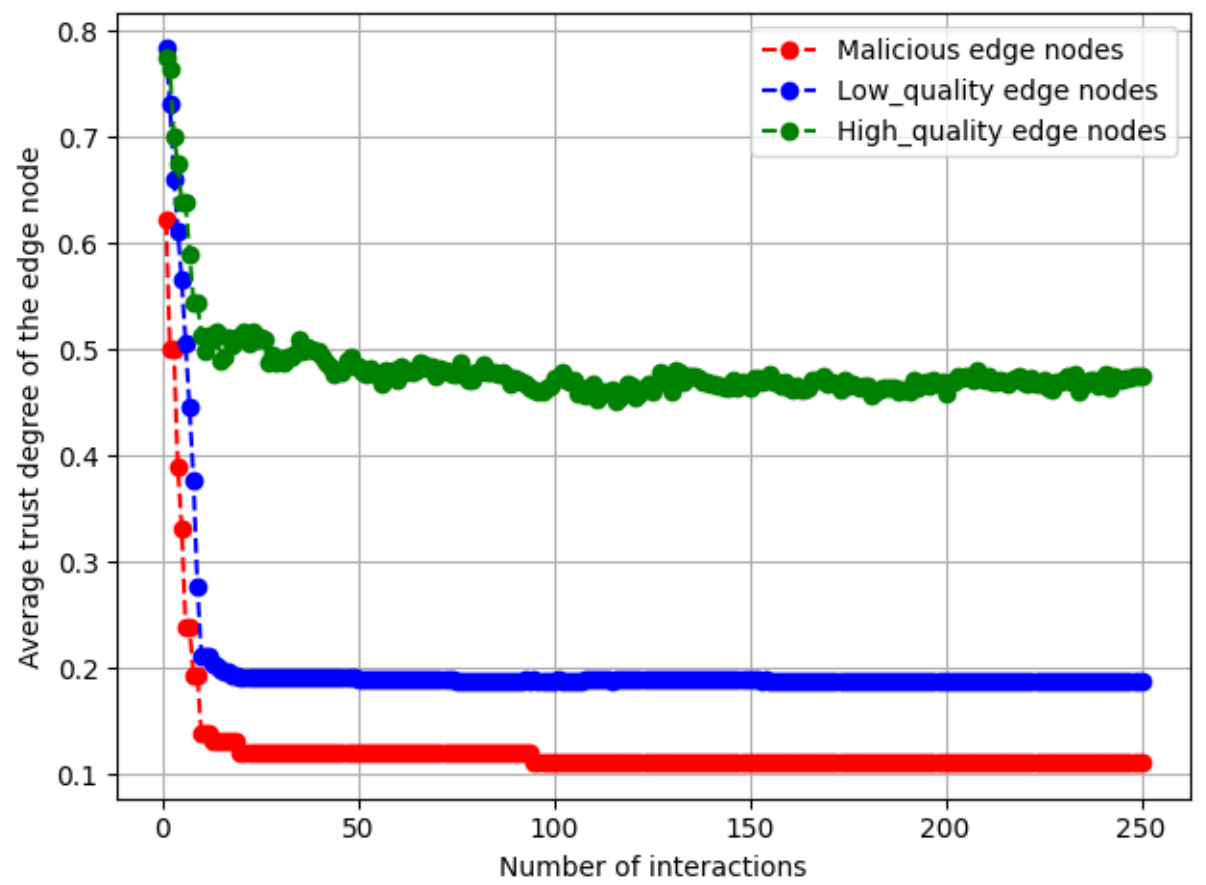

**FIGURE 26.** The edge nodes trust level with 30 Mobile Users

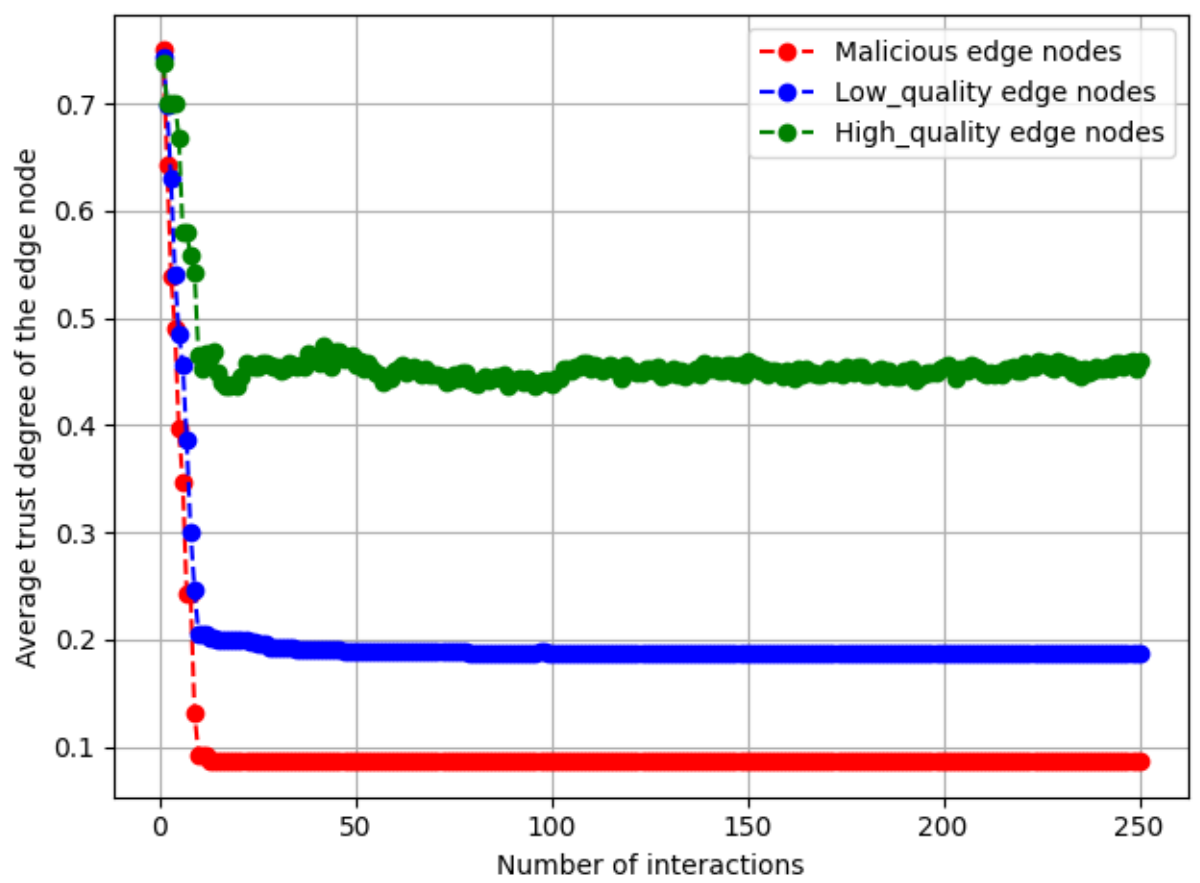

**FIGURE 27.** The edge nodes trust level with 40 Mobile Users

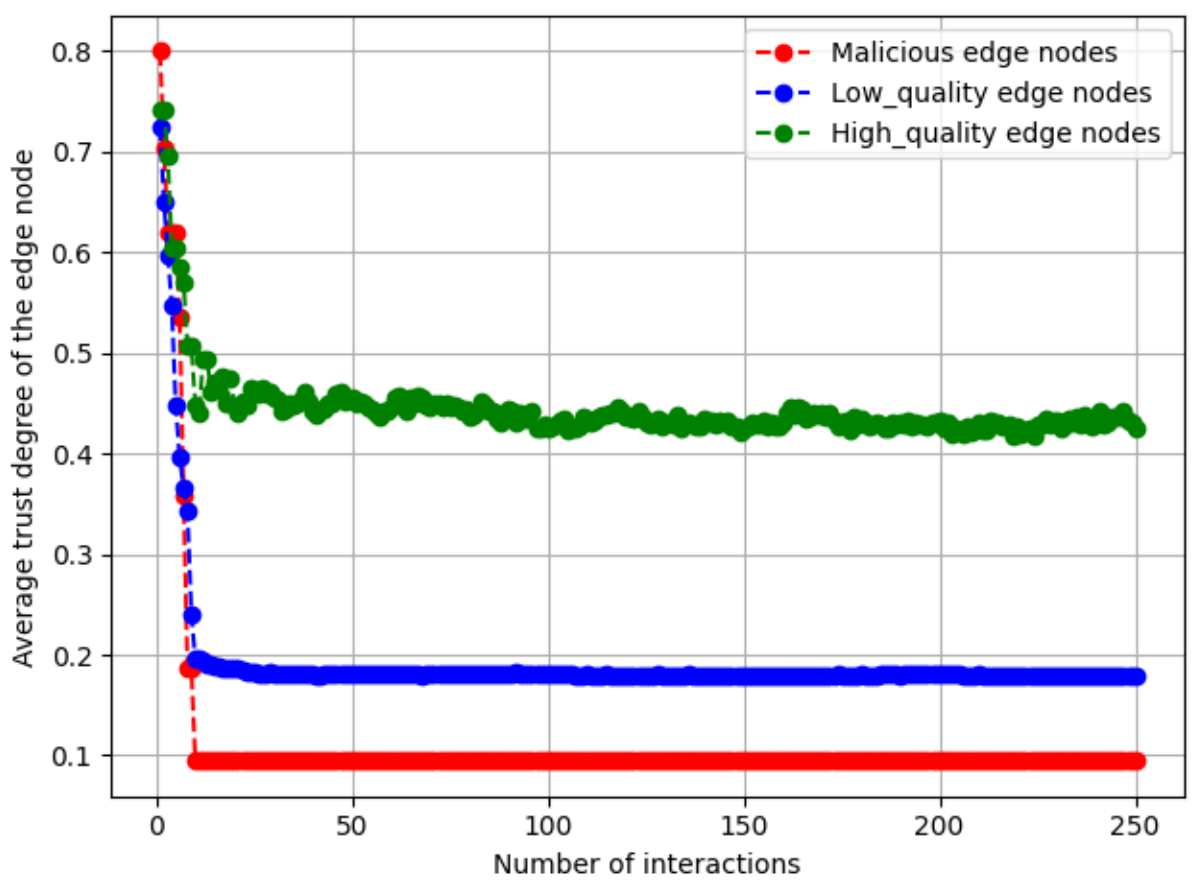

**FIGURE 28.** The edge nodes trust level with 50 Mobile Users.

### 6) ROBUSTNESS TO NOISE IN USER FEEDBACK

In this section, network noise is considered solely in relation to the satisfaction evaluation provided by the Mobile User (MU) regarding the quality of services offered by the Edge Server (ES). The trust computation algorithm, implemented as a smart contract on the blockchain, is immutable and cannot be tampered with. Similarly, the trust data stored on the blockchain is secure and unalterable. However, the input to this contract—namely, the Mobile User's evaluation of the Edge Server's service quality ($E_{ji}$)—is collected off-chain and may be affected by network noise before being submitted and permanently recorded on the blockchain. Consequently, noise can cause an inaccurate evaluation of service quality to be registered, which in turn affects the trust score assigned to the ES. Therefore, this paper examines the robustness and accuracy of the proposed model under varying network noise conditions and evaluates its performance using confusion-matrix-based metrics.

Figures 29 and 30 illustrate the effects of increasing noise in the data—specifically, in the MU's evaluation of the service quality provided by the ES (as per Equation 13)—on the confusion matrix metrics. This experiment involved 20 ESs and 30 MUs, with the ESs distributed as 20% malicious, 60% low-quality, and 20% high-quality.

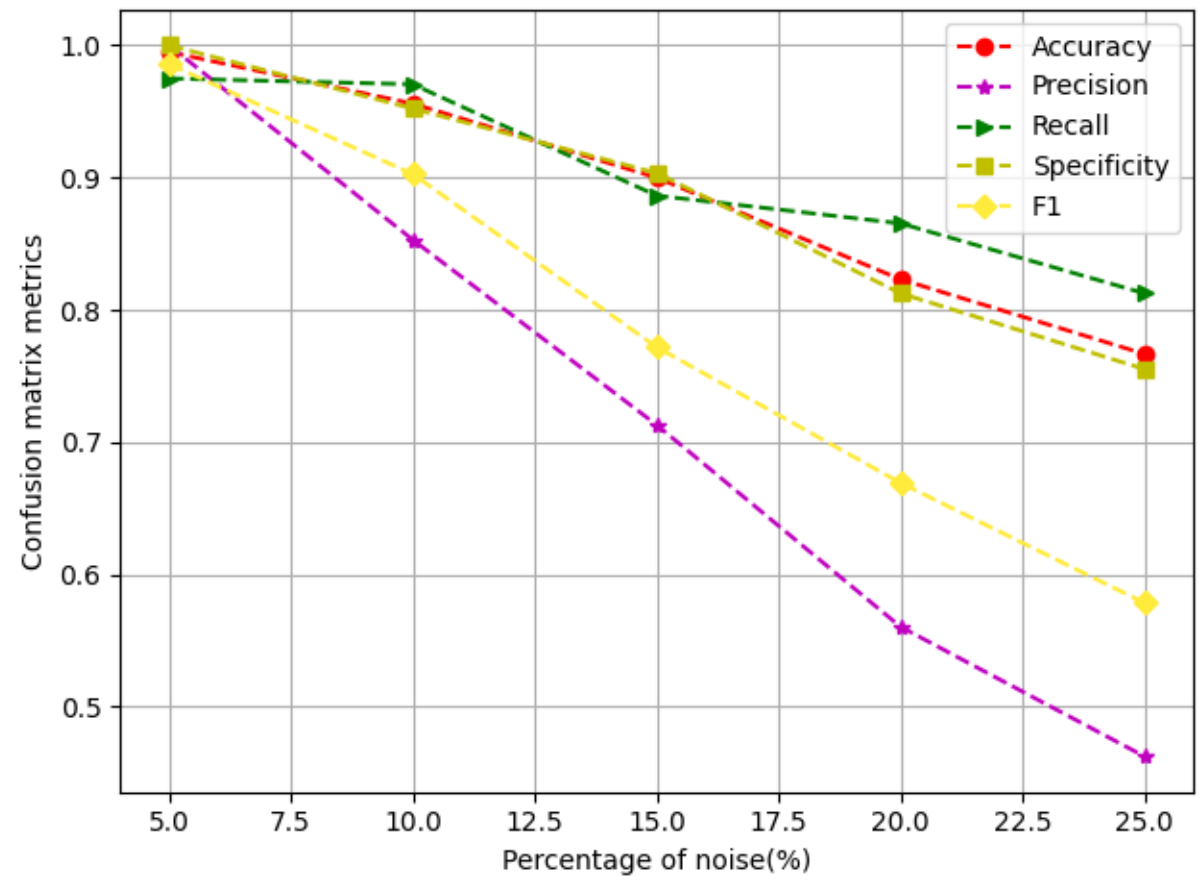

**FIGURE 29.** Macro Average of Confusion Matrix Metrics for Two Classes

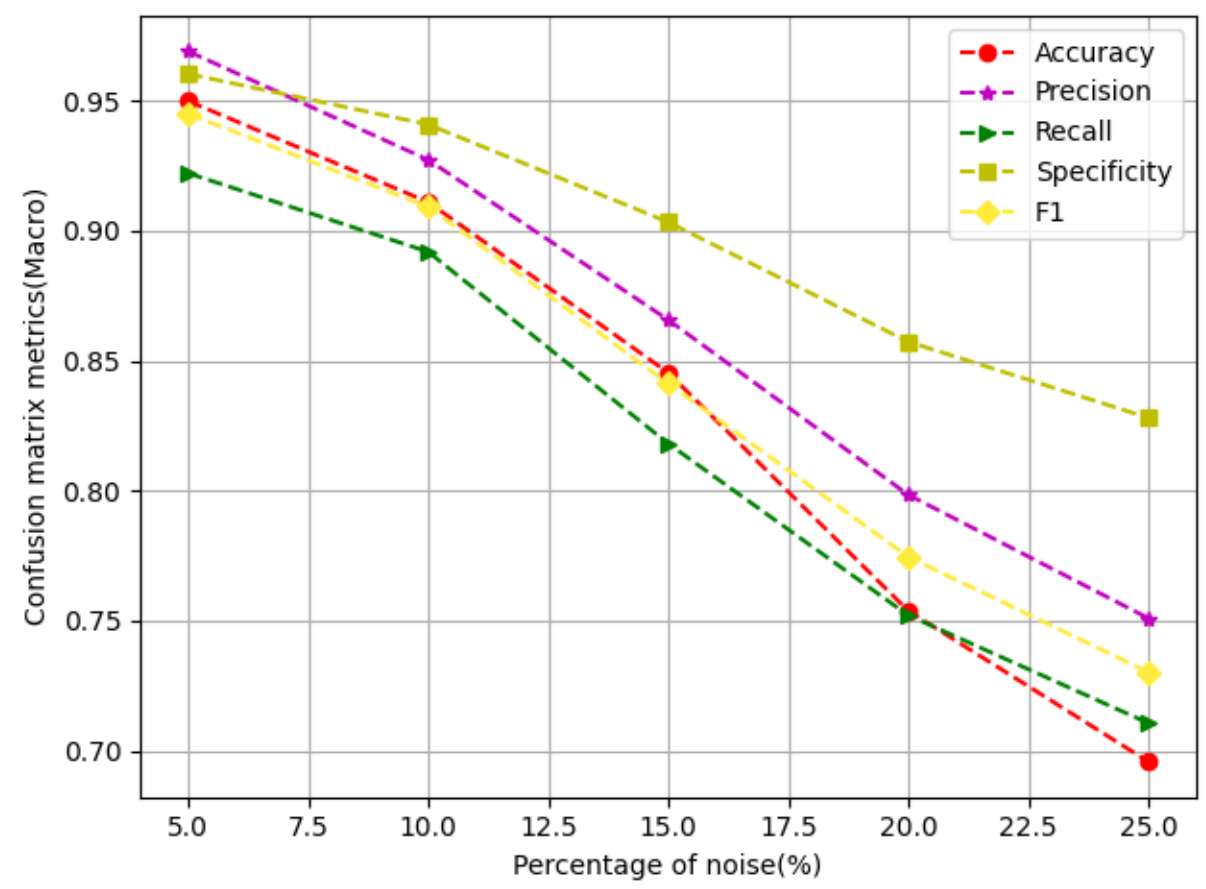

**FIGURE 30.** Macro Average of Confusion Matrix Metrics for Three Classes

In Figure 29, the ESs are divided into two categories: honest and malicious, with high-quality and low-quality ESs classified as honest. The specificity and recall values remain above 0.9 until the noise level reaches 10%.

However, when noise exceeds 10%, malicious servers begin to be incorrectly identified as honest.

Figure 30 shows the macro-average of the confusion matrix metrics across the three categories: malicious, low-quality, and high-quality servers. Here, the specificity and recall values also hover around 0.9 until noise reaches 10%. Beyond this threshold, there is an increasing likelihood of misclassifying malicious ESs as either low-quality or high-quality, and vice versa.

While our model's evaluation stands independently on metrics like accuracy and recall, direct comparisons with alternative methods were not pursued here, as many prior approaches do not account for the unique security threats inherent to edge caching, such as collusion among mobile nodes, identity forgery, and content manipulation.

However, further evaluation under real-world network conditions and with comparison to baseline approaches could provide deeper insights into the generalizability and performance of the proposed model. Overall, the results indicate that the proposed model accurately classifies edge servers and manages trust effectively across different scenarios, including variations in server quality and user demand. The method also demonstrates resilience to noise and manipulative behavior, making it a suitable solution for secure and efficient edge caching environments.

### C. Security Model Evaluation and Attack Resilience

This section evaluates the resilience of the proposed model against common attacks in edge computing environments. Each type of attack represents a potential threat to the trust and integrity of edge servers, especially in decentralized and mobile settings where malicious nodes may attempt to exploit weaknesses in trust management mechanisms. The proposed model effectively counters these threats using techniques such as smart contract-based calculations and multi-layered authentication protocols, contributing to the security and stability of the edge network.

#### 1) SELF-PROMOTION ATTACKS

In these attacks, a node may inflate its importance by providing positive recommendations for itself to be selected as a server, only to later stop providing services or deliver faulty ones [34]. Within the proposed approach, such an attack occurs when an ES manipulates its trust score to appear more trustworthy, subsequently providing malicious services. However, in the proposed method, trust is calculated within a smart contract and securely stored on the blockchain, preventing the ES from altering the trust calculation algorithm or independently manipulating its trust score. Additionally, collusion among groups of MUs to artificially increase a server's trust is mitigated by the credibility calculation of MU evaluations in Equation (9). The inclusion of the correction parameter $f$ in the direct trust calculation (Equation 13) further ensures that malicious ESs cannot easily boost their trust score through a small number of successful interactions. Instead, an ES can only achieve higher trust by consistently delivering high-quality services.

#### 2) BAD MOUTHING ATTACK

In this common attack against trust management systems, malicious nodes spread false negative information about honest nodes to damage their reputation [11]. Specifically, a malicious MU could attempt to reduce the trust of reliable ESs by reporting low satisfaction with their services, thereby decreasing the ES's trust score. However, in the proposed method, the impact of such manipulation by a malicious MU is minimized due to the calculation of mobile users' evaluation credibility in Equation (9) and the indirect trust calculation in Equation (14). Since the weight of indirect trust in the trust calculation formula (Equation 15) is set at 0.9, indirect trust plays a more significant role in the overall trust assessment of the ES, thereby reducing the effect of any false reports from malicious MUs.

#### 3) WHITEWASHING ATTACKS

Whitewashing attacks occur when a node with a poor reputation attempts to reset its reputation by re-entering the network under a new identity [14]. Both mobile users (MUs) and edge servers (ESs) are potential perpetrators of this type of attack. The proposed method's resistance to whitewashing attacks is analyzed in the following scenarios:

- **Attack by Edge Server:** In this scenario, an edge server might change its identity to hide its previous trust level and rejoin the network to continue offering services. The proposed approach mitigates this threat through stringent authentication mechanisms managed by the authentication server. Since edge servers must authenticate themselves before participating in the network, they cannot alter their identity or reset their trust level within the cloud.
- **Attack by Mobile User:** This scenario involves a mobile user changing their identity to either provide or receive services under a new identity, effectively resetting their reputation. The proposed method addresses this issue by implementing robust authentication protocols, both by the authentication server and at the edge server level. These dual layers of authentication ensure that MUs cannot change their identity when moving between servers, thereby preventing them from circumventing the system and resetting their reputation.

Unlike conventional models that fail to enforce persistent identity verification, our approach ensures traceable user identities through blockchain-based authentication, effectively preventing whitewashing attacks.

#### 4) ON-OFF ATTACKS

In an On-Off attack, a node alternates between honest (ON) and malicious (OFF) behavior. During the ON phase, the node builds up trust, which it later exploits in the OFF phase to attack the network [35]. In the proposed approach, a malicious edge server (ES) may randomly provide good and bad services. During its good phase, it gains trust, which it then uses to carry out malicious activities. To counter this, the proposed method incorporates a correction parameter $f$ in Equation (13), which prevents a malicious ES from quickly building up high trust with just a few successful interactions. This mechanism ensures that only those ESs consistently providing high-quality services can achieve a higher trust level.

In contrast to baseline models that accumulate trust rapidly from a few successful interactions, our model uses the correction parameter f to ensure that trust increases only with sustained service quality, thereby reducing vulnerability to on-off behavioral patterns.

#### 5) COLLISION ATTACK

In a Collision Attack, several malicious nodes collaborate to artificially increase or decrease the trust value of other edge computing nodes [11]. The proposed method neutralizes this threat by calculating the credibility of each mobile user's (MU) evaluation as part of the trust calculation process (as outlined in Equation 9). This mechanism prevents MUs from colluding to manipulate the trust of an ES, whether the goal is to inflate or deflate its trust level.

#### 6) EDGE SERVER ATTACKS

Edge Server Attacks occur when malicious ESs compromise data integrity or disrupt network services [11]. In the proposed method, if an ES engages in malicious activities, its trust level diminishes over time. Eventually, the trust mechanism will lead to the removal of such an ES from the network, thereby preserving the overall integrity and security of the system.

While prior models removed untrusted edge servers, they often lacked clear classification and removal rules. Our model uses a dynamic trust threshold, periodically updated, to identify servers falling below the threshold. These servers are automatically removed from the network, ensuring fairness and operational precision.

#### 7) BALLOT STUFFING ATTACKS

Ballot Stuffing Attacks involve an attacker inflating the reputation of malicious nodes by providing positive recommendations, thereby increasing their chances of being selected as service providers [36]. In this scenario, MUs may falsely report high satisfaction with services from malicious ESs, aiming to boost these servers' trust levels. The proposed method prevents such manipulation by using smart contracts to calculate ES trust and storing the results on the blockchain, ensuring the integrity of trust calculations. Additionally, the credibility of MU evaluations, as calculated in Equation (9), further mitigates the risk of collusion aimed at artificially enhancing the trust of an ES. Since indirect trust is given greater weight in the overall trust calculation, attempts to manipulate trust through direct trust alone have minimal impact.

Unlike conventional schemes that treat all feedback equally, our method evaluates the credibility of each MU’s input before incorporating it into the trust score, thereby limiting the impact of collusion-based reputation inflation.

Compared to prior models that often overlook the reliability of feedback sources or rely solely on direct trust accumulation, the proposed method offers a more comprehensive defense framework by incorporating credibility-based evaluation, the corrective parameter $f$, and immutable smart contract enforcement.

Overall, the proposed method demonstrates strong capability in distinguishing between malicious and high-quality edge servers, even under challenging conditions such as data noise and mobile node collusion. This performance highlights its suitability for dynamic, resource-constrained edge environments, where reliable content caching is critical.

Collectively, these design elements distinguish our approach from prior models, which often lack dynamic weighting and credibility assessment, rendering them more susceptible to manipulation.

### *D. Performance Evaluation*

#### 1) COMPUTATIONAL AND COMMUNICATION COSTS

Given the nature and characteristics of IoT-based networks, analyzing their efficiency is essential [37]. To evaluate the practical performance of the proposed two-stage consensus mechanism in such environments, we conducted a series of empirical tests measuring the following parameters.

- CPU Usage (%): The average processor load during trust aggregation and block validation.
- Memory Usage (MB): The memory consumed per node during consensus operations.
- Transaction Latency (sec): The average time required for a transaction to be confirmed and added to the blockchain.

The experiments were repeated 50 times. The consensus mechanism ran on a virtual machine with a 4-core Intel Xeon E5-2620 v4 CPU at 2.10 GHz and 8 GB of RAM. All results were averaged across the 50 rounds, using feedback data that included 10% noise.

The observed average values are shown in Table 11.

TABLE 11
COMPUTATIONAL AND COMMUNICATION COSTS FOR PROPOSED APPROACH

| Metric | Measured Value |
|---|---|
| CPU Usage (%) | 28.6% |
| Memory Usage (MB) | 62.4 MB |
| Transaction Latency (sec) | 0.42 sec |

#### 2) PERFORMANCE COMPARISION

To further assess the effectiveness of the proposed trust model, we conducted a comparative analysis with two recent blockchain-based trust management schemes: the edge caching trust model by Xu et al. [10], and the vehicular trust framework proposed by Zhang et al. [15]. The comparison focuses on two key performance metrics under similar experimental settings: detection accuracy and trust evaluation latency.

As shown in Table 12, our model achieves higher detection accuracy and faster trust convergence latency under noisy environments compared to the referenced works. Furthermore, only our model explicitly incorporates collusion resistance through credibility filtering and indirect trust mechanisms. These results confirm the proposed scheme's robustness and practical value in real-world IoT environments.

TABLE 12
PERFORMANCE COMPARISON

| Model | Detection Accuracy (%) | Latency (sec) | Noise Level | Collusion Defense |
|---|---|---|---|---|
| Proposed Model | 92.3 | 0.49 | 10% | ✓ |
| Xu et al. (2020) [10] | 86.7 | 0.58 | 10% | ✗ |
| Zhang et al. (2021) [14] | 91.2 | 0.47 | 10% | ✓ |

### *E. Sensitivity Analysis*

To examine the robustness of the proposed trust model against parameter variations, we conducted a sensitivity analysis on two key hyperparameters: the initial trust threshold ($\varrho$) and the weight coefficient ($\alpha$) used in the trust update function.

#### 1) SENSITIVITY TO INITIAL TRUST THRESHOLD

The parameter $\varrho$ controls the baseline trust assigned to newly joined nodes. A very low value may reduce the influence of new nodes, while a high value may allow malicious nodes to gain trust too quickly. To study its impact, we varied $\varrho$ in the range of {0.05, 0.1, 0.2, 0.3, 0.4}, and evaluated the resulting trust computation accuracy, false positive rate (FPR), and trust convergence time under a fixed noise level of 10%. The results are shown in Table 13.

TABLE 13
SENSITIVITY ANALYSIS WITH RESPECT TO INITIAL TRUST THRESHOLD

| $\varrho$ Value | Detection Accuracy (%) | False Positive Rate (%) | Convergence Time (sec) |
|---|---|---|---|
| 0.05 | 91.2 | 4.8 | 0.61 |
| **0.10** | **94.3** | **3.7** | **0.42** |
| 0.20 | 93.5 | 5.2 | 0.47 |
| 0.30 | 91.9 | 6.1 | 0.54 |
| 0.40 | 89.7 | 8.3 | 0.63 |

The results indicate that the model performs optimally when $\varrho$ is around 0.1. Lower values reduce responsiveness, while higher values introduce instability and increase false trust.

#### 2) SENSITIVITY TO TRUST UPDATE WEIGHT

The parameter $\alpha$ determines the weight given to new evidence versus historical trust. We tested $\alpha$ values in {0.05, 0.1, 0.2, 0.3, 0.5}. As shown in Table 14, performance remains relatively stable in the range [0.05, 0.2], with optimal results at $\alpha = 0.1$. Beyond this range, trust values fluctuate excessively and reduce stability.

TABLE 14
SENSITIVITY ANALYSIS WITH RESPECT TO TRUST UPDATE WEIGHT

| $\alpha$ Value | Detection Accuracy (%) | Trust Value Variance | Convergence Time (sec) |
|---|---|---|---|
| 0.05 | 92.8 | 0.011 | 0.48 |
| **0.10** | **94.3** | **0.008** | **0.42** |
| 0.20 | 93.1 | 0.015 | 0.47 |
| 0.30 | 91.5 | 0.023 | 0.52 |
| 0.50 | 88.9 | 0.037 | 0.60 |

## VI. Conclusion and Future Work

This paper presents a blockchain-based trust management system that enables reliable selection of edge servers by evaluating both direct and indirect trust, helping identify and eliminate malicious servers in the network. Trustworthiness is determined based on mobile users' satisfaction, with trust data securely stored on the blockchain through smart contracts. Edge servers are classified into high-quality, low-quality, and malicious categories, allowing users to connect with servers that provide secure and reliable services. The approach incorporates an authentication scheme to prevent identity spoofing and a two-stage consensus mechanism designed to address the mobility and energy constraints of mobile devices. Additionally, a reward system encourages mobile users to share their cache capacity when edge servers experience shortages, with incentives provided in proportion to service quality, promoting optimal resource sharing.

Future research will expand this framework by implementing a two-layer blockchain structure to enhance scalability and minimize transaction delays. A lightweight private blockchain will handle frequent trust calculations, while trust data will periodically be registered on a global blockchain via smart contracts, balancing privacy and efficiency. Further, Quality of Service (QoS) parameters will be integrated into the trust model to accommodate users' diverse service needs and account for varying server performance, ultimately reinforcing the robustness and adaptability of this trust management approach in dynamic edge computing environments. Additionally, we aim to develop a cache resource allocation algorithm for edge servers with a time complexity lower than $O(n^2)$ and improved scalability.

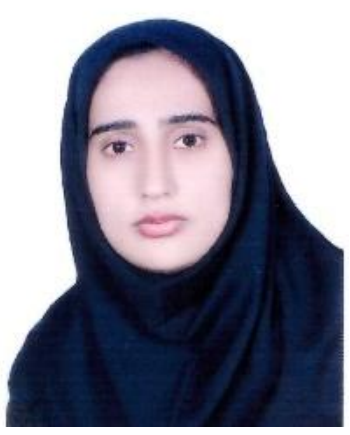

**Motahare Ebrahimi** completed her B.Sc. in Information Technology from 2012 to 2016. She then pursued her M.Sc. in Computer Engineering with a specialization in Computer Networks from 2020 to 2023 at Yazd University, Yazd, Iran. Her interests include web design and conducting research in the fields of networking and security.

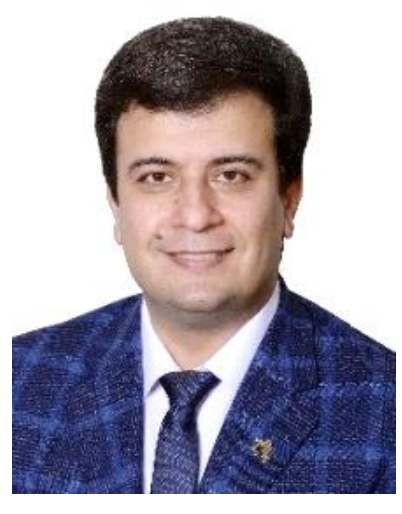

**Nastooh Taheri Javan** (*Senior Member, IEEE*) is an Assistant Professor with the Computer Engineering department at Imam Khomeini International University (IKIU), Qazvin, IRAN. Dr. Taheri Javan was post-doctoral fellow at Amirkabir University of Technology (Tehran Polytechnic), Tehran, IRAN, where he completed his M.S. and Ph.D. in computer engineering in 2007 and 2017, respectively. His research interests lie in the area of wireless computer networks and network coding theory, spanning from theory to design and implementation. Dr. Taheri Javan has actively collaborated with researchers in various disciplines of computer science, particularly resource management on problems at the network architecture area.

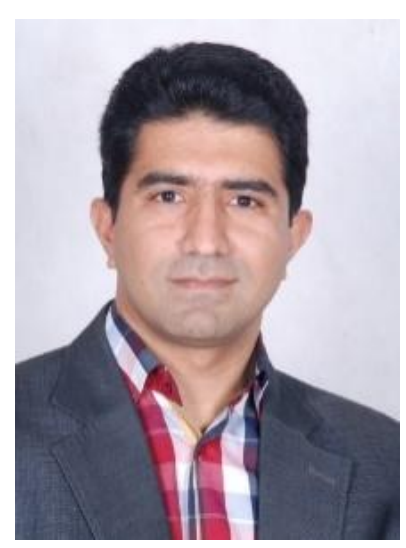

**Seyedakbar Mostafavi** is an Associate Professor in the Department of Computer Engineering at Yazd University, Iran. He holds a B.Sc. in Information Technology from Sharif University of Technology and a Ph.D. in Computer Networks from Amirkabir University of Technology (Tehran Polytechnic). Dr. Mostafavi leads the "Information Technology Enterprise Architecture" research lab at Yazd University. His research focuses on resource management in cloud computing and wireless networks, making him a valuable asset to the field of computer engineering. Dr. Mostafavi actively contributes to the academic community by frequently reviewing papers for international journals and conferences.

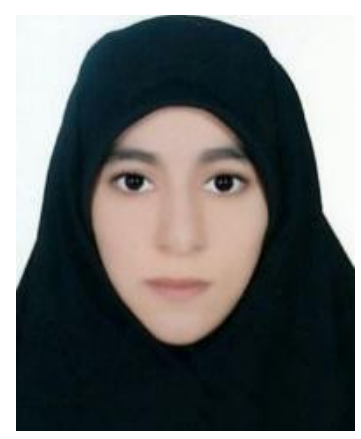

**Fatemeh Pakzaban** earned her B.Sc. in Computer Engineering from Yazd University, Iran, in 2020. She continued her studies at the same institution, obtaining her M.Sc. degree in Computer Networking in 2024. Her primary research interests lie in the fields of Network Security and the Internet of Things.